\documentclass[10pt, aps,prc,twocolumn,superscriptaddress,preprintnumbers,
amsmath, 
floatfix,
longbibliography,
nofootinbib
]{revtex4-1}
\usepackage[T1]{fontenc}
\usepackage[utf8x]{inputenc} 
\usepackage{adjustbox}          
\usepackage[caption=false]{subfig}
\usepackage{url}
\usepackage{color}
\usepackage{float}
\usepackage[pdftex,colorlinks=true, linkcolor = blue, citecolor=blue,urlcolor=blue, bookmarksnumbered=true, bookmarksopen=true]{hyperref}
\usepackage{longtable}
\usepackage{amsfonts}
\usepackage{dsfont}
\usepackage{wrapfig,bm} 
\usepackage[normalem]{ulem}
\usepackage{MnSymbol}
\usepackage{float}

\usepackage{xcolor}

\newcommand{\beq}{\begin{equation}}
\newcommand{\eeq}{\end{equation}}
\newcommand{\bea}{\begin{eqnarray}}
\newcommand{\eea}{\end{eqnarray}}

\begin{document}
\title{Measures of complexity and entanglement in many-fermion systems}
\author{Aurel Bulgac}%
\affiliation{Department of Physics,%
  University of Washington, Seattle, Washington 98195--1560, USA}
\author{Matthew Kafker}%
\affiliation{Department of Physics,%
  University of Washington, Seattle, Washington 98195--1560, USA} 
\author{Ibrahim Abdurrahman}%
\affiliation{Department of Physics,%
  University of Washington, Seattle, Washington 98195--1560, USA}

\date{\today}

\begin{abstract}

There is no unique and widely accepted definition of the complexity measure  (CM) of a many-fermion wave function in the presence of interactions. 
The simplest  many-fermion wave function is a  Slater determinant. In shell-model or configuration interaction (CI) 
and other related methods, the state is represented as a superposition of a large number of Slater determinants, which in case 
of CI calculations reaches about 20 billion terms~\cite{Johnson:2018}. Although in practice this number has been used as a CM for decades,  it is ill defined: it is not unique, and
it depends on the particular type and the number of single-particle wave functions used to construct the Slater determinants.

The canonical wave functions/natural orbitals~\cite{Lowdin:1956a,Lowdin:1956,Bardeen:1957,Bogoliubov:1958,Valatin:1958,Gennes:1999,Ring:2004} 
and their corresponding occupation probabilities are intrinsic properties of any 
many-body wave function, irrespective of the representation, and they provide a unique solution to characterize the CM. 
The non-negative orbital entanglement entropy, which vanishes for a Slater determinant, provides the simplest CM, while
a more complete measure of complexity is the entanglement spectrum.  We illustrate these aspects in the case of 
a complex non-equilibrium time-dependent process, induced nuclear fission described within a 
real-time Density Functional Theory framework extended to superfluid systems, which can 
describe simultaneously the long-range and the short range correlations between fermions. 

The orbital entanglement entropy of the fissioning 
nucleus illustrates the localization mechanism of the many-body wave-function in Fock and/or Hilbert space. 
The (minimal) number of Slater determinants required to represent such a complex many-body wave function
with a well defined number of particles in the case presented here is 
about $10^{500}$. The realistic  case of the highly non-equilibrium nuclear fission process illustrated here is equivalent to a 
system of $23.328\times10^9$ interacting quantum spin-1/2 particles, likely the largest system where 
quantum entanglement has been studied so far. 
 
\end{abstract}

\preprint{NT@UW-22-02}

\maketitle  

\section{Introduction}

Only two years after \textcite{Schrodinger:1926} published his equation,
the representation of the wave functions for an interacting many-body systems became a question of major concern. 
For a system of $N$ spinless particles in 3D such a wave function would require $(n_s^3)^{N-1}$ complex numbers, where $n_s$ is the number of 
discrete spatial points in 1D. The smallest spatial separation between two spatial lattice points $l$ 
determines the maximum momentum cutoff $\Lambda=\pi\hbar/l$. 
The simplest solution suggested 
almost a century ago for a system of many fermions was to use the Hartree-Fock (HF) 
approximation~\cite{Hartree:1928,Slater:1928,Slater:1930,Fock:1930} and its later extension 
the Hartree-Fock-Bogoliubov (HFB) approximation~\cite{Bardeen:1957,Bogoliubov:1958,Valatin:1958,Gennes:1999,Ring:2004}, 
which in 3D and in the absence of any symmetry, is still numerically challenge even nowadays.  
At the same time, short-range correlations (SRCs) are critical in obtaining {\it ab initio} accurate 
descriptions of many-fermion systems. In the case of the dynamics of many-fermion systems out of equilibrium this is still a question 
which  awaits to be fully addressed microscopically~\cite{Bulgac:2022}.

For fermion systems the elementary building 
block in constructing a many-fermion wave function is the Slater determinant, also known as the HF wave-function. In the presence of pairing correlations the generalized 
Slater determinant, the HFB many-body wave function, plays a similar role, and often in the case of nuclei it
requires a particle projection.
Despite the short-range and strong character of the interactions between nucleons, many single-particle and 
collective properties of nuclei can be calculated quite successfully using  mean-field theory approaches such as  HF and HFB, 
Landau-Migdal theory for Fermi liquids, shell-model calculations, etc. In all these approaches a rather limited number of single-particle orbitals  are typically used.
When pairing correlations are present  the Fermi surface is diffuse and the number of single particle orbitals needed in order to describe the nuclear masses and the 
low-energy excited states in most applications is at most about twice as large as the total particle number~\cite{Bohr:1969,Ring:2004}.

For decades in nuclear physics the correlations have typically been treated with rather low momentum cutoffs $\Lambda$ and 
their effects have been rolled into effective (not always synonymous with accurate) mostly phenomenological  low-energy interactions. 
The typical argument used in calculations of open-shell nuclei was that the energy of the ground state converged 
quite rapidly as a function of the chosen cutoff. This is expected in the case of a variational approach, since errors of order ${\cal O}(\delta)$ 
in the many-body wave-function lead to errors ${\cal O}(\delta^2)$ in the energy near the minimum. 
\textcite{Anderson:1984}, in discussing the treatment of electronic systems, 
characterized this kind of situation as   the ``Quantum Chemists' Fallacy No. 1 and 2,''
of which even Wigner was partially guilty, as ``you may get 
pretty good energetics out of a qualitatively wrong state.''
The perfect example is the case of a superconductor, in which despite the fact 
that the contribution to the ground state energy from the condensation energy is 
practically negligible, meaning the ground state energy can be evaluated with 
sufficient accuracy in the absence of pairing correlations, the 
wave-function with pairing correlations leads however to qualitative changes,
which otherwise  would have been completely overlooked. 

The Slater determinants form  complete set of $N$-particle many-body states. The number of Slater determinants in an expansion of a many-fermion 
wave function has a very strong dependence on the size of the single-particle (HF) basis set. In CI or  shell-model 
calculations~\cite{Johnson:2018}, which are used to construct the ground-state and a few excited states only, 
the dimension of the Hamiltonian matrix to be diagonalized, before any symmetry restrictions are imposed,  is
\begin{align} 
\frac{N_{sp}!}{N!(N_{sp}-N)!} \approx \left ( \frac{N_{sp}}{N}\right)^N \; \text{typically} \ll (n_s^3)^{N-1},\label{eq:orb}
\end{align}
where  $N_{sp}$ is the size of the adopted single-particle Hilbert space. Notice that $N_{sp}$ is the only adjustable parameter
in Eq.~\eqref{eq:orb} for a given particle number $N$.
Similar arguments apply also for other many-body techniques such as coupled cluster approaches, the
generator coordinate method,  and 
certain implementations of the Quantum Monte Carlo method. 
Does this number accurately describe the complexity of a CI many-body wave function? 
In the case of time-dependent processes it is well known that the dynamics are governed in general by statistical factors, namely 
by the number of accessible states available for the system to evolve into for ergodic systems. 
As a result, underestimating the number of basis states could lead to inaccurate results, while overestimating 
this number can lead to unnecessary calculations. However, equilibration times might be significantly longer 
than specific characteristic times, in which case statistical arguments are not applicable.  

The main questions we address in this work are: 
i) Does a useful measure or measures of the complexity of a many-body wave function exist?
ii) Does a minimal set of $N_{sp}$ single-particle states exist and what are its properties? 
This question received an answer many years ago~\cite{Lowdin:1956a,Lowdin:1956,Coleman:1963,Coleman:1963a,Davidson:1972} 
iii) How can one construct such a basis set easily?  iv) Does a measure or measures of complexity of a many-body wave function 
shed any new light on non-equilibrium processes, where (local) thermalization has not had enough time to occur? 
 
More than 70 years ago \textcite{Levinger:1951} invoked the quasi-deuteron model and the short range-character of the proton-neutron correlations
in order to describe the nuclear photo-effect. Using realistic nucleon interactions various authors  observed that very high momentum single-particle 
states are occupied with significant probability~\cite{Benhar:1993,Pandharipande:1997}. As  in the case of Levinger's quasi deuteron model, 
the presence of the SRCs, reflected in the significant single-particle probabilities of high-momentum states, were crucial  in 
order to describe the results of the $(e,e'p)$ experiments~\cite{Quint:1986,Quint:1987}, 
which showed that deep single-particle levels were occupied with an unexpectedly 
low probability~\cite{Pandharipande:1997,Aumann:2021} $n_{sp} \approx 0.6$. 
Brueckner's framework was a parallel approach favored for decades~\cite{Day:1967,Negele:1982} used
to include, although not explicitly, the role of SRCs into the mean field treatment nuclei.
 
There is a rather wide range of observables, which 
cannot be reproduced accurately in calculations in mean field type of treatments.  One example is  the nucleon  
momentum distribution, which has been studied theoretically and experimentally in cold atom and nuclear systems
for a long  time~\cite{Quint:1986,Quint:1987,Sartor:1980,Frankfurt:1981,Frankfurt:1988, Benhar:1993,Ciofi-degli-Atti:1996,Pandharipande:1997,Piasetzky:2006,Sargsian:2005,Schiavilla:2007,Sargsian:2014,Hen:2017,
Zwerger:2011,Braaten:2011,Castin:2011,Anderson:2015,Porter:2017,Stewart:2010,Hen:2014,Hen:2017,Aumann:2021,Cruz-Torres:2021}, 
confirming  Shina Tan's~\cite{Tan:2008a,Tan:2008b,Tan:2008c} prediction made for systems with zero-range interactions 
 that the high-momentum distribution  behavior $n_k=C/k^4$ is in fact a
generic feature of strongly interacting many-fermion systems, and thus a qualitative and quantitative 
feature of such systems both in and out of equilibrium. 
The important conclusion of many studies of nuclear systems was that approximately 20\% of 
the spectral strength is found for momenta $k>k_F$. 

In current studies of the masses and low energy spectra of nuclei the role of SRCs is captured in a reduced space of single-particle orbitals 
using renormalization group techniques~\cite{Bogner:2010,Furnstahl:2013}, and the SRCs never appear explicitly.  
Tropiano {\it et al.}~\cite{Tropiano:2021,Tropiano:2022}
demonstrated recently how Levinger's quasi deuteron model and the effect of SRCs on the nucleon momentum distribution
can be reproduced at a low momentum resolution using the similarity renormalization group (SRG) approach~\cite{Bogner:2010,Furnstahl:2013}. 
Within the SRG approach the scattering properties and energy spectra of very light nuclei are reproduced with impressive accuracy, although  the quality  
degrades with increasing atomic mass~\cite{Hu:2022a}, likely due to a simpler theoretical treatment of SRCs at only the NNLO level~\cite{Maris:2021}. 
However, the simplification provided by the SRG approach in calculating ground state 
and low-energy excited state properties results
in a rather complex and opaque structure of various observables, 
in particular the nucleon momentum distribution, which become complex 
many-body operators with the result that  ``\emph{the best choice of scale may not be so clear for 
analyzing SRC experiments because of the tradeoffs.}~\cite{Tropiano:2021}'' 
The extension of the SRG approach to time-dependent non-equilibrium phenomena is yet to be realized. The dynamics are 
controlled by gain and loss in case of kinetic equations, and therefore by the availability of final states in particular. Then, it is not obvious 
whether that SRG approach, which operates in a limited single-particle space, could describe time-dependent phenomena, 
such as fission or many-nucleon transfer reactions in heavy-ion collisions.
Many outstanding issues, concerning the relevance of the SRCs in low-energy nuclear physics in particular,
remaining to be addressed in the near future were discussed in the recent papers, see
\emph{Nuclear Forces for Precision Nuclear Physics: A Collection of Perspectives}~\cite{Tews:2022} and \emph{Dense Nuclear Matter Equation of State from Heavy-Ion Collisions}~\cite{Sorensen:2023}. As \textcite{Miller:2020} notes, the three scales relevant to 
low-energy nuclear physics, size of nuclei, average separation between nucleons, and the nucleon size are basically of the same order of magnitude, and thus there is effectively no scale separation and the effects of SRCs can be measured~\cite{Miller:2019}.

In sections~\ref{section:II} and \ref{section:III} 
we describe the general properties of the canonical wave functions, which are needed to evaluate 
the canonical occupation probabilities and the orbital entanglement entropy,
which provide measures of the complexity and of the many-body wave functions. The complexity of a many-body wave-function 
can be characterized by the degree of non-similarity to a Slater determinant. 
Canonical wave functions were introduced a long time ago, in connection with describing pairing correlations,
as the eigenvectors of the one-body density matrix. 
The mathematical framework for describing superfluid fermionic systems was formulated 
in terms of quasiparticles by \textcite{Bogoliubov:1958} and \textcite{Valatin:1958}. \textcite{Zumino:1962} 
and \textcite{Bloch:1962} have shown that one can introduce a particular set of quasiparticles, with similar properties 
to the  set used by \textcite{Bardeen:1957} (BCS), the canonical set of states, see also Ref.~\cite{Ring:2004}. 
\textcite{Lowdin:1956a,Lowdin:1956} introduced the natural spin orbitals as the eigenvectors of the one-body density matrix. 
The definitions of the canonical wave functions and the natural orbitals are mathematically identical, 
yet have been used in different 
contexts, particularly extensively in chemistry~\cite{Davidson:1972}, but lately also in nuclear 
physics~\cite{Pandharipande:1984,Stoitsov:1993,Reinhard:1999,Dobaczewski:1996,Tajima:2004,
Tichai:2019,Robin:2021,Hoppe:2021,Fasano:2022,Chen:2022,Hu:2022,Hagen:2022,Kortelainen:2022,Tichai:2022}, 
often without realizing that they represent the same complete set of orthonormal single-particle orbitals,
namely the canonical set of states. 

It has been proven that if one 
intends to represent a correlated many-body wave-function as a sum over 
Slater determinants, the natural orbitals, or in another words, the canonical 
wave functions set is the optimal set, namely the smallest size single-particle basis 
set~\cite{Lowdin:1956a,Lowdin:1956,Davidson:1972,Coleman:1963,Coleman:1963a}. 
Mathematically it is obvious that the canonical wave functions or the 
natural orbitals form a full orthonormal set, but as far as we can judge from the literature 
many properties of this set were  never studied, as only a small reduced number of 
canonical wave functions was ever extracted numerically and only some properties 
of this reduced set were discussed. We show here that the canonical wave 
functions have some very distinctive, even striking  and peculiar 
properties, which were never discussed in literature.  

We demonstrate that the canonical wave functions are basically of three types: i) a subset similar to usual mean field single-particle wave functions; 
ii) a subset of wave functions corresponding to occupation probabilities $n_k\approx C/k^4$, oscillate much faster than the mean field type of 
single-particle wave-functions,  are fully localized inside the system, have rather small occupation probabilities, and 
are typically ignored in evaluation of the ground state properties of nuclei; 
iii) a subset of canonical wave functions localized outside the system and which play an insignificant role in defining physical
properties of the system. The first subset has a size comparable to the particle number. 
The size of the second subset, not explicitly discussed in literature,  is typically  
an order of magnitude larger in size than the first subset (or even larger for small spatial resolutions) and its size is 
determined by the level of spatial resolution adopted or the momentum cutoff
\begin{align} 
\Lambda=\frac{\pi\hbar}{dx}, 
\end{align}
which is defined by the adopted spatial resolution $dx=l$.
Only canonical wave functions of  type i) and ii) are relevant in order to accurately evaluate 
properties of a many-body system and in particular SRCs  and the entanglement or the 
Boltzmann and Shannon entropies. We will show that the combined
number of states  of type i) and ii) is approximately given by phase space volume
\begin{align}
g\frac{4\pi}{3} r_0^3A \frac{4\pi}{3} \Lambda^3 \frac{1}{ (2\pi\hbar)^3 }=\frac{8\pi^2}{9}\left( \frac{r_0}{dx}\right)^3 A,
\end{align} 
where $g=4$ is the spin-isospin degeneracy factor and $r_0 =1.2$ fm for nuclei. 
In Section~\ref{section:IV} we discuss the 
definition of the orbital entropy for a system of indistinguishable particles.

In section~\ref{section:V}  we illustrate the insight the time evolution of the entanglement or Boltzmann entropy can provide 
in the case of quantum non-equilibrium processes, specifically induced nuclear fission, and demonstrate that the entanglement 
entropy and therefore the size of the physically relevant canonical set has a non-monotonic time-dependence, of similar 
origin as the widely discussed many-body localization in 1D systems \cite{Milburn:1997,Vidal:2003,Korepin:2004,Kitaev:2006,Levin:2006,Li:2008,
Chuchem:2010,Pal:2010,Bardarson:2012,Cohen:2016,Abanin:2019,Sinha:2020,Wimberger:2021, Liu:2022,Mueller:2022,Schneider:2022}. 
In the dynamics of isolated systems the evolution of the entanglement entropy plays the role of thermodynamic entropy 
for local observables~\cite{Calabrese:2005,Calabrese:2006,Alba:2017}. 
The manner the case of indicted fission is described theoretically is 
similar to what in condensed matter literature is called quenching, when a system is prepared as the stationary state of a nuclear Hamiltonian 
subject external constraints  and then it is evolved in time under a pure Hamiltonian with no constraints. 
The realistic case of the highly non-equilibrium nuclear fission process illustrated here is equivalent to a system of 
$8\times(N_xN_yN_z)^2=8\times(30^2\times60)^2=23.328\times10^9$ interacting quantum
spin-1/2 particles. This nuclear system is  likely by orders of magnitude 
the largest system where quantum entanglement has been studied so  far in literature.

The experience gathered during more than a decade of 
studying non-equilibrium processes in nuclear systems and cold-atom systems seem to point towards a rather unexpected 
emerging scenario. It was demonstrated that pairing-like correlations can emerge at very large excitations energies, when they are not supposed to   
exist~\cite{Bulgac:2011,Bulgac:2017,Bulgac:2019c,Bulgac:2020,Magierski:2022}. One can partially understand such behavior using the semiclassical 
Nordheim~\cite{Nordheim:1928} and Boltzmann-Uehling-Uehlenbeck~\cite{Uehling:1933} approach to quantum kinetic phenomena, which 
has been recently extended in a pure quantum framework~\cite{Bulgac:2022}. The neutron-neutron and proton-proton 
collisions are captured in a time-dependent extension of the Density Functional Theory (DFT)~\cite{Bulgac:2013a,Bulgac:2019c} 
by including the pairing field. As we will demonstrate here, as well in the case of induced fission, see Ref.~\cite{Bulgac:2022} 
and section~\ref{section:V}, the time-dependent pairing fields lead to a large population of the high-momentum 
single-particle states, a process which is expected in non-equilibrium phenomena, as systems typically evolve towards large regions of allowed phase-space.  
We present our conclusions in the last section~\ref{section:VI}.

\section{The canonical basis or natural orbital set}\label{section:II}

We show here how using the canonical~\cite{Ring:2004} (introduced for treating superfluid systems) 
or natural orbital~\cite{Lowdin:1956a,Lowdin:1956} basis set one can get insight into how many single-particle 
states are  needed to accurately describe various properties of a physical system. 

As the single-particle strength is spread by interactions over 
a wide energy interval the structure of the many-body wave function is always very complex. Even in a reduced single-particle Hilbert space, as 
used in shell-model calculations, the number of contributing Slater determinants is of the order of tens of billions~\cite{Johnson:2018}, 
a number which depends very strongly on the type of the single-particle set of wave functions used. This number is only optimal if one uses the 
canonical or natural orbital set~\cite{Lowdin:1956a,Lowdin:1956,Davidson:1972,Coleman:1963,Coleman:1963a}.
The complexity can be quantified for any quantum state $|\Phi\rangle $ by evaluating the 
orbital entanglement/quantum Boltzmann one-body 
entropy~\cite{Nordheim:1928,Uehling:1933,Bulgac:2022,Klich:2006,Amico:2008,
Horodecki:2009,Haque:2009,Eisert:2010,Boguslawski:2014,Gigena:2015,Robin:2021}
\begin{align} 
S = &- g\sumint_k n_k\ln n_k  \nonumber\\
&- g\sumint_k  [1-n_k]\ln[1-n_k],\label{eq:ent}
\end{align}
where $g$ is the spin-isospin degeneracy, $\sumint$ implies summation over discrete and integration over continuous 
variables respectively, and $n_k$ are the canonical occupation probabilities
\begin{align} 
&\int d\zeta \, n(\xi,\zeta)\phi_k(\zeta) = n_k\phi_k(\xi),\, 0\le n_k\le 1,\label{eq:nnn1} \\ 
& \sumint_\xi \phi_k^*(\xi) \phi_l(\xi)=\delta_{kl}, \label{eq:ort1}\\ 
&N = \sumint_k n_k ,
\end{align}
where $N$ is the total particle number and $n(\xi,\zeta)$ is the number density matrix defined accordingly~\footnote{Some authors prefer the definition of the density matrix normalized to 1~\cite{Coleman:1963,Bengtsson:2017}, instead particle number $N$, as in this case the space of density matrices becomes convex.}
\begin{align}   
&n(\xi,\zeta) = \langle \Phi|\psi^\dagger(\zeta)\psi(\xi)|\Phi\rangle.  \label{eq:ntr1}
\end{align}
Here $\psi^\dagger (\xi)$ and $ \psi (\xi)$ are the field operators 
for the creation and annihilation of a particle with coordinate $\xi=({\bm r},\sigma,\tau)$  (spatial, spin, and isospin coordinates) and $|\Phi\rangle$ 
is an arbitrary quantum many-body state, either static or time-dependent. The wave function $\Phi$ can describe either a static or 
time-dependent many-body system, and therefore the canonical occupation numbers and the corresponding canonical wave function can be 
time-dependent as well.

The many-body wave-function  $\Phi$ can be a member of a Hilbert space, if the particle number 
is well defined, or of the Fock space, in which case it will contain components with different particle numbers. We will discuss here both cases 
in the context of nuclear fission. The one-body density $n(\xi,\zeta)$ can also be defined as~\cite{Lowdin:1956a,Lowdin:1956,Coleman:1963,Davidson:1972} 
\begin{align} 
&\!\!\!\!n_N(\xi_1,...,\xi_N,\zeta_1,...,\zeta_N)=\Phi(\xi_1,...,\xi_N)\Phi^*(\zeta_1,...,\zeta_N),\\
&\!\!\!\!n(\xi,\zeta) = N\! \int\prod_{k=2}^Nd\xi_kn_N(\xi,\xi_2,...,\xi_N,\zeta,\xi_2,...,\xi_N).
\end{align}

The orbital entanglement entropy $S$ defined in Eq.~\eqref{eq:ent} is non-vanishing 
in the ground state of any interacting system~\cite{Srednicki:1993,Klich:2006,Boguslawski:2014,Robin:2021,Bulgac:2022}, 
unlike the textbook thermodynamic entropy. 
The orbital entanglement entropy $S$ attains its minimum value in the case of a pure Slater determinant $S_{min}=0$, when $ n_k\equiv 1$ or $n_k\equiv 0$,   
and its maximum value when $n_k \equiv N/N_{sp}$, where $N_{sp}$ is the dimension of the single-particle space and 
the single-particle strength is spread uniformly over the entire Hilbert space.
The entropy $S$, which is thus a measure of the complexity of the many-body wave-function, 
can be evaluated accurately only when very high-momentum occupation probabilities 
up to values $n_k \approx 10^{-6}$ are taken into account, see section~\ref{section:V}. 

From a quantum information science (QIS) point of view it is convenient to use the Shannon definition of the entropy~\cite{Bengtsson:2017}, 
and use instead a rescaled set of canonical occupation probabilities (typically arranged in decreasing order)
\begin{align}
&\tilde{n}_k=\frac{n_k}{N},\quad \sumint_k \tilde{n}_k = 1,\quad 0\le \tilde{n}_k\le \frac{1}{N},\\
&\overline{S} =   -\sumint_k \tilde{n}_k \log_2 \tilde{n}_k = -\frac{1}{N}\sumint_k n_k\log_2 n_k +\log_2 N. \label{eq:Shannon}
\end{align}
In the case of Fermi systems the minimum and maximum possible values of the Shannon entropy are
\begin{align}
&\overline{S}_{min} = \log_2 N, \quad \overline{S}_{max} =\log_2 {N_{sp}},
\end{align} 
The minimum value for $\overline{S}$ is achieved only in the case of a Slater determinant for $N > 1$ particles, and for any 
superposition of Slater determinants, $\overline{S}> \log_2 N$.  

Both entropies $S$  and $\overline{S}$  obviously characterize the level of complexity of the many-body wave-function:  
the extent to which particle 
interactions spread the single-particle strength over the entire spectrum. It is important to notice that both entropies can attain their
minimum values for states, which as a rule do not correspond to a minimum total energy.   Basically both of these entropies characterize, 
in slightly different manners, the degree of the entanglement of the many-body system and  
henceforth we will use only the Boltzmann entropy $S$ from this point onwards.

\section{Properties of the canonical wave functions}\label{section:III}

The canonical states or the  natural orbitals $\phi_k(\xi)$ form a  
complete set 
\begin{align} 
\sumint_k \, \phi_k(\xi)\phi_k^*(\zeta) = \delta(\xi-\zeta). \label{eq:full}
\end{align}
Since $n(\xi,\zeta)$ basically vanishes when either spatial coordinate is well outside the system, 
any function $f(\xi)$ with support outside the system is automatically an eigenstate of $n(\xi,\zeta)$ with $n_k\approx 0$.
The canonical states in the case of a finite nucleus in vacuum form a set with cardinality $\mathfrak{c}$, the 
cardinality of $\mathbb{R}^3$. If the nucleus is simulated in a finite box then the number of canonical states
is countable and the set has the cardinality $\aleph_0$, the cardinality of the integers. 
Since for a stable nucleus 
the number density decays exponentially at large distances, the description of a bound nucleus in a sufficiently large 
simulation box should be sufficient, and a smaller set of single-particle wave functions with cardinality $\aleph_0$ should always suffice.
 
 An eigenfunction $\phi_k(\xi)$ with $n_k >0$ has its support largely inside the system, where the support  
of $n(\xi,\zeta)$ is, and one can presume that it oscillates with the maximum momentum
$p_\text{max} =\sqrt{2m|U|} $ that a typical nuclear mean field 
can support for a bound state, where $U\approx -50$ MeV is the depth of the mean field. 
One can then conjecture that the total number
of states with $n_k>0$ is of the order of the total number of bound  quantum states a nuclear mean field can sustain
\begin{align}  
N_\text{max} \approx \frac{ 4\pi p_\text{max}^3}{3} \times \frac{4\pi r_0^3 A}{3}\times  \frac{1}{(2\pi\hbar)^3}\approx  0.5A  , \label{eq:Nmax}
\end{align} 
where $r_0 = 1.2$ fm and $A$ is the total number of nucleons, and where we did not account for spin and isospin degrees of freedom. 
(The spin-isospin degeneracy was not accounted for here.) Since both 
normal number and anomalous densities are constructed from canonical qpwfs, with 
strictly non-vanishing occupation probabilities $0<n_k\le 1$,  it then follows that  only a finite  
set of such functions  is likely needed to represent the densities. 
We will show below that Eq.~\eqref{eq:Nmax} grossly underestimates the size of the canonical basis set with $n_k>0$.
Using Eqs.~\eqref{eq:ntr} and (\ref{eq:nnn1}) it follows that the density matrix $\overline{n}(\xi,\zeta)$ has the same eigenfunctions $\phi_k(\xi)$
\begin{align}
\int d\zeta\, \overline{n}(\xi,\zeta) \phi_k(\zeta) = (1-n_k)\phi_k(\xi). \label{eq:non-nnn}
\end{align}
This equation may be used to construct the canonical states 
localized mostly outside the system.

One can introduce the time-reversal canonical orbitals~\cite{Bohr:1969}, not necessarily identical to those defined in Eq.~\eqref{eq:kbar},
\begin{align}
\phi_{\overline k}(\xi) = i\sigma_y\phi_k^*(\xi), \label{eq:tr}
\end{align}
where $\sigma_y$ is the Pauli matrix.
\begin{figure} 
\includegraphics[width=1.0\columnwidth]{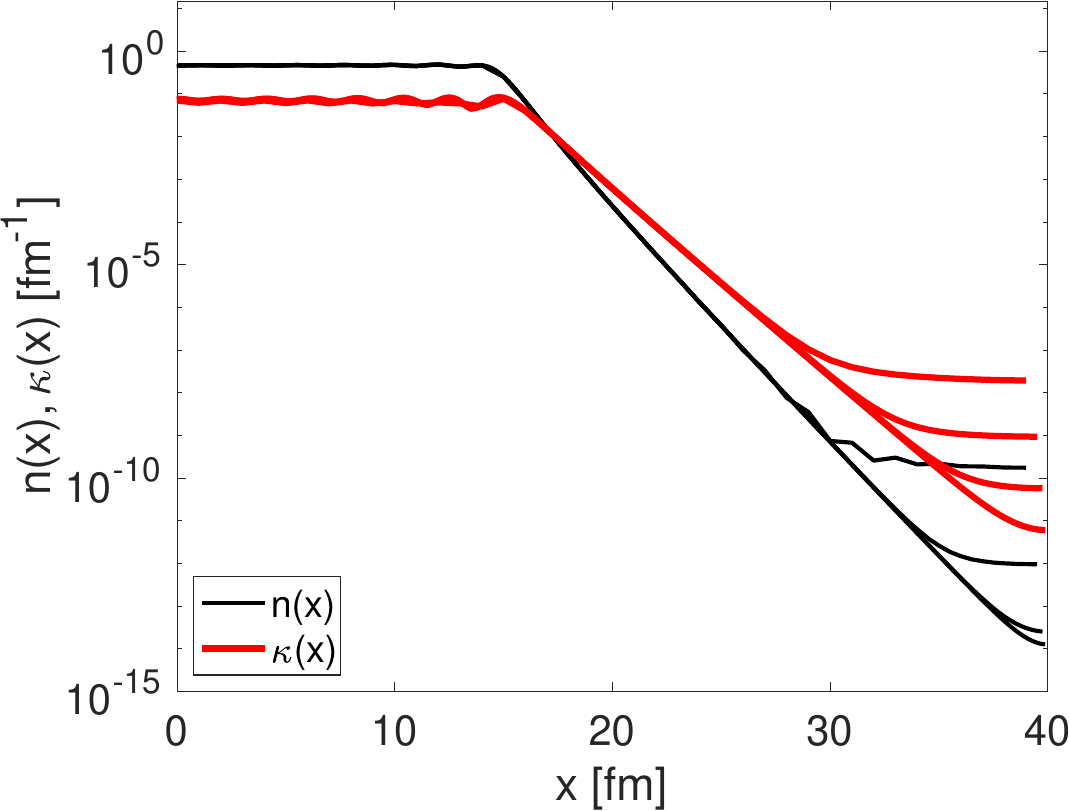}
\caption{ \label{fig:number}  
The normal number (black)  and anomalous (red) densities for $x\ge 0$,  for four lattice constants in decreasing order  
$dx=1, 0.5, 0.25, 0.125$ fm. $n(x)$  and $\kappa(x)$ stand for $n(x,x)$ and $\kappa(x,x)$ respectively. }
\end{figure}  

\begin{figure}
\includegraphics[width=1.0\columnwidth]{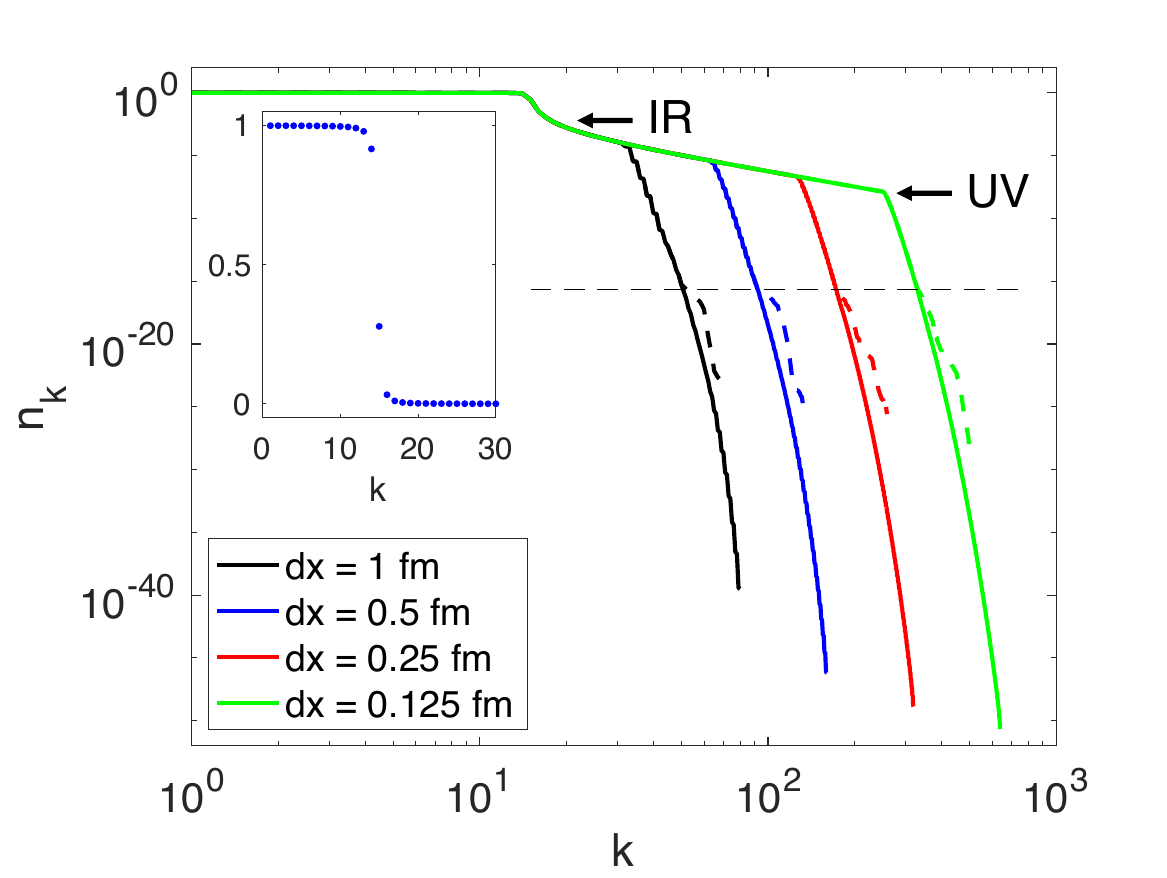}
\caption{ \label{fig:nk}   
The canonical occupation probabilities for four lattice constants $dx$ in a log-log scale, corresponding to 
momentum cutoff $\Lambda = \pi\hbar/dx$. 
In the inset we plot $n_k$ in the linear scale close to the Fermi surface.  The results obtained with increased 
machine precision $10^{-40}$ are shown with continuous solid lines. 
The dashed horizontal black line shows the level of typical machine double precision $10^{-16}$ and 
dashed lines for $n_k$ show the corresponding results obtained for the occupation probabilities.}
\end{figure}

We will first illustrate the properties of the canonical wave functions with some generic numerical results obtained for a 1-dimensional example, 
which  retains all the qualitative features of a 3-dimensional system. For the sake of simplicity we have chosen
a 1-dimensional system with potential and pairing fields
\begin{align}
V(x) &= \frac{V_0}{1+\frac{\cosh(x/a)}{\cosh(R/a)}},\\
\Delta(x) &= \frac{\Delta_0}{1+\frac{\cosh(x/a)}{\cosh(R/a)}},
\end{align}
where we will use the notation for the spatial coordinate $-\infty< x< \infty$, $V_0 = -50$ MeV, $\Delta_0= 3$ MeV, $R=r_0A^{1/3}= 14.9$ fm, $a=0.5$ fm, 
and $\mu = -5$ MeV.  (We avoid using a Woods-Saxon potential well in order not to generate singularities of 
the derivatives of the wave functions at the origin, which would lead to unphysical long momentum tails of the wave functions.)   
We solved the non-self-consistent  SLDA or HFB
equations for the qpwfs~\cite{Bulgac:2007,Bulgac:2020}, using the Discrete Variable Representation (DVR) method~\cite{Bulgac:2013}
\begin{align} 
\begin{pmatrix} 
H-\mu  & \Delta \\
\Delta  & -H +\mu
\end{pmatrix}
\begin{pmatrix}
{\textrm u}_{k} \\
{\textrm v}_{k}
\end{pmatrix}
= E_k
\begin{pmatrix}
{\textrm u}_{k} \\
{\textrm v}_{k}
\end{pmatrix}\label{eq:tdslda} 
\end{align} 
in a box of size $L=80$ fm and with four different lattice constants $dx =1, 0.5, 0.25, 0.125$ fm,  
where 
\begin{align}
H=-\frac{\hbar^2}{2m} \frac{d^2}{dx^2}+V(x)
\end{align}
and $m$ is the nucleon mass, in the absence of spin-orbit interaction. 
Eq.~\eqref{eq:tdslda} are for the components ${\textrm u}_k(x)$ with spin-up 
and ${\textrm v}_k(x)$ with spin-down. The equations for the  components ${\textrm u}_k(x)$ with spin-down 
and ${\textrm v}_k(x)$ with spin-up are obtained from these equations by changing 
the sign of the pairing field $\Delta(x)$ only~\cite{Bulgac:2007,Bulgac:2020}. 
The SLDA equations for cold fermionic gases and nuclei have the same structure in this case.
It is straightforward to extend this type of analysis to more complicated geometries, for example the pasta phase in neutron star crusts, 
or the superconductor-normal metal-superconductor (SNS) junctions in condensed matter physics. The case discussed here is equivalent 
to a NSN junction. This analysis equally applies to infinite periodic systems. 

\begin{figure}
\includegraphics[width=1.0\columnwidth]{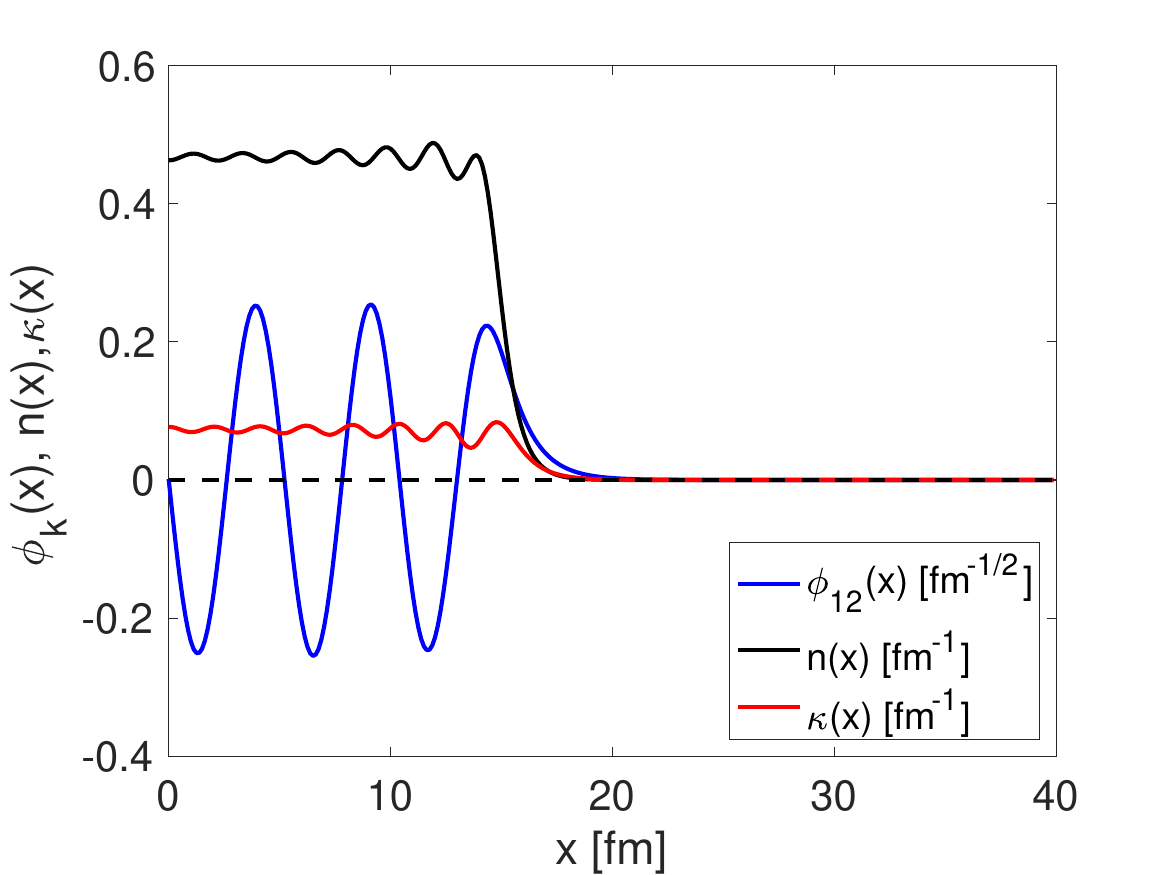}  
\caption{ \label{fig:phi_12}  
The canonical wave-function $\phi_{12}(x)$ and occupation probability $n_{12}=0.978$ along with profiles of 
the number density $n(x)$ and of the anomalous density $\kappa(x)$ in the case $dx=0.125$ fm. 
Since $V(x)=V(-x)$ and $\Delta(x)=\Delta(-x)$, all these functions have well defined spatial parities and we represent these quantities only for $x\ge 0$. }
\end{figure}  

\begin{figure}
\includegraphics[width=1.0\columnwidth]{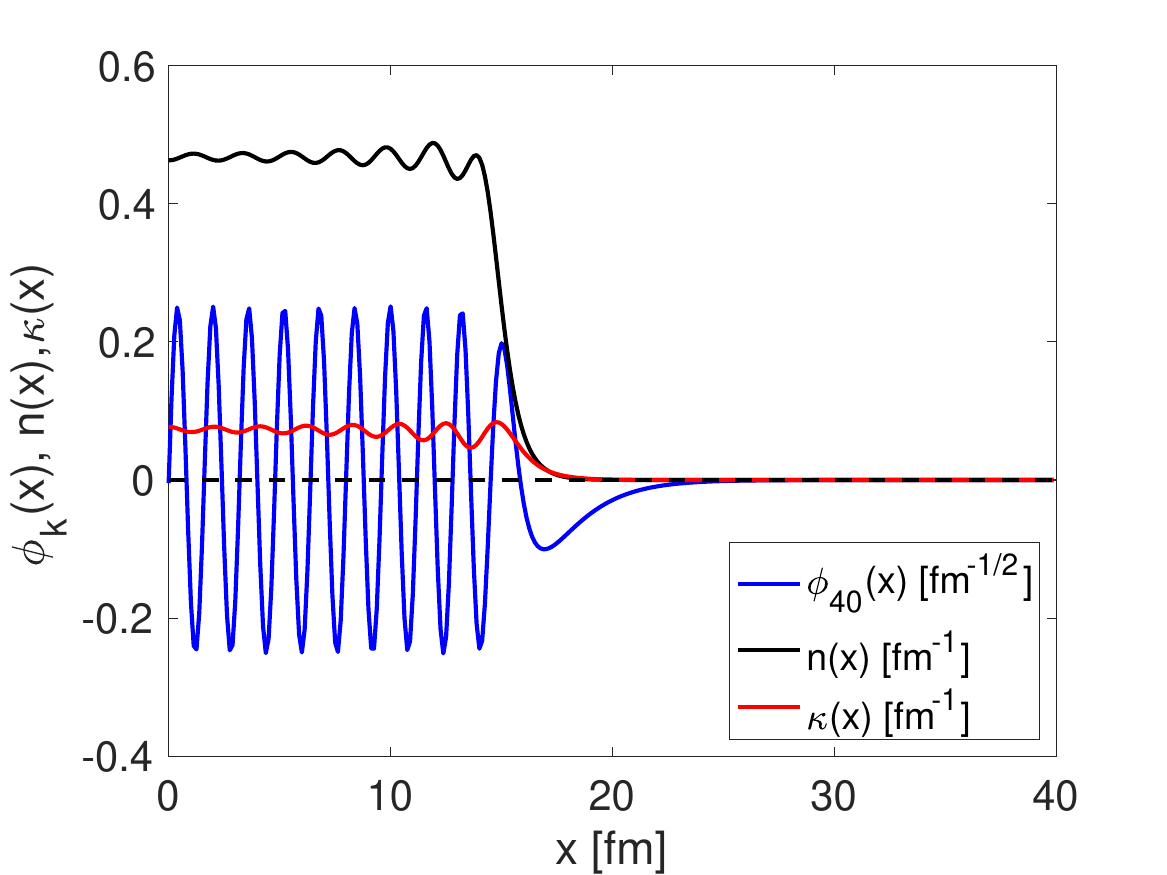}
\caption{ \label{fig:phi_44}  
The same as Fig.~\ref{fig:phi_12} for $\phi_{40}(x)$ and $n_{40} = 1.298\times 10^{-4}$, 
localized inside the system, with $k$ in the interval between the IR-knee and the UV-knee.} 
\end{figure}  

This 1-dimensional model is equivalent to solving the SLDA equations for a spherical system, in this case for $s$-orbitals,
with orbitals  $\phi_k(x)=-\phi_k(-x)$ and $x\ge 0$ in the present formulation. For a 3-dimensional spherical system the wave functions would be 
$\psi({\bm r}) = \phi(r)/r Y_{lm}(\theta,\phi)$ and $r=x\ge 0$. For angular momenta $l>0$ one has to add the centrifugal potential $\hbar^2l(l+1)/2mr^2$. 
In the presence of the centrifugal barrier  a classically forbidden region appears near the origin and some of the corresponding canonical wave functions 
for $l>1$ will have the character of ``exterior'' functions with occupation probabilities $n_k$ beyond the UV-knee shown in Fig.~\ref{fig:nk}.
The 1-dimensional normal number density $n(x)$  here is only for the fermions with spin-down, which in the case of even fermion particle number 
is identical to the normal number density of the spin-up particles. 
As shown in Ref.~\cite{Bulgac:1980} the anomalous density $\kappa(x)$ has longer exponential tails than the number density $n(x)$. 
This longer tail of the pairing field becomes particularly important as one approaches the nucleon drip-line.
This behavior should be also apparent in the profiles of $V(x)$ and $\Delta(x)$, an aspect which we neglected here and which does not 
change the qualitative behavior of these densities, see Fig.~\ref{fig:number}.  Fig.~\ref{fig:number} also shows that 
with increasing spatial resolution ($dx\rightarrow 0$) the normal density is 
more accurately reproduced at larger distances. We have also checked that Eqs.~\eqref{eq:norm} and (\ref{eq:anom}) in the Appendix
correctly reproduce the normal and anomalous densities when using the canonical wave functions.  From the spatial behavior of the canonical wave-functions 
illustrated in Figs.~\ref{fig:phi_12} and \ref{fig:phi_44} it is obvious that they can be obtained with real accuracy using semi-classical quantization 
conditions, as they are almost perfect stationary standing waves in an almost perfect square well potential.

\begin{figure}
\includegraphics[width=1.0\columnwidth]{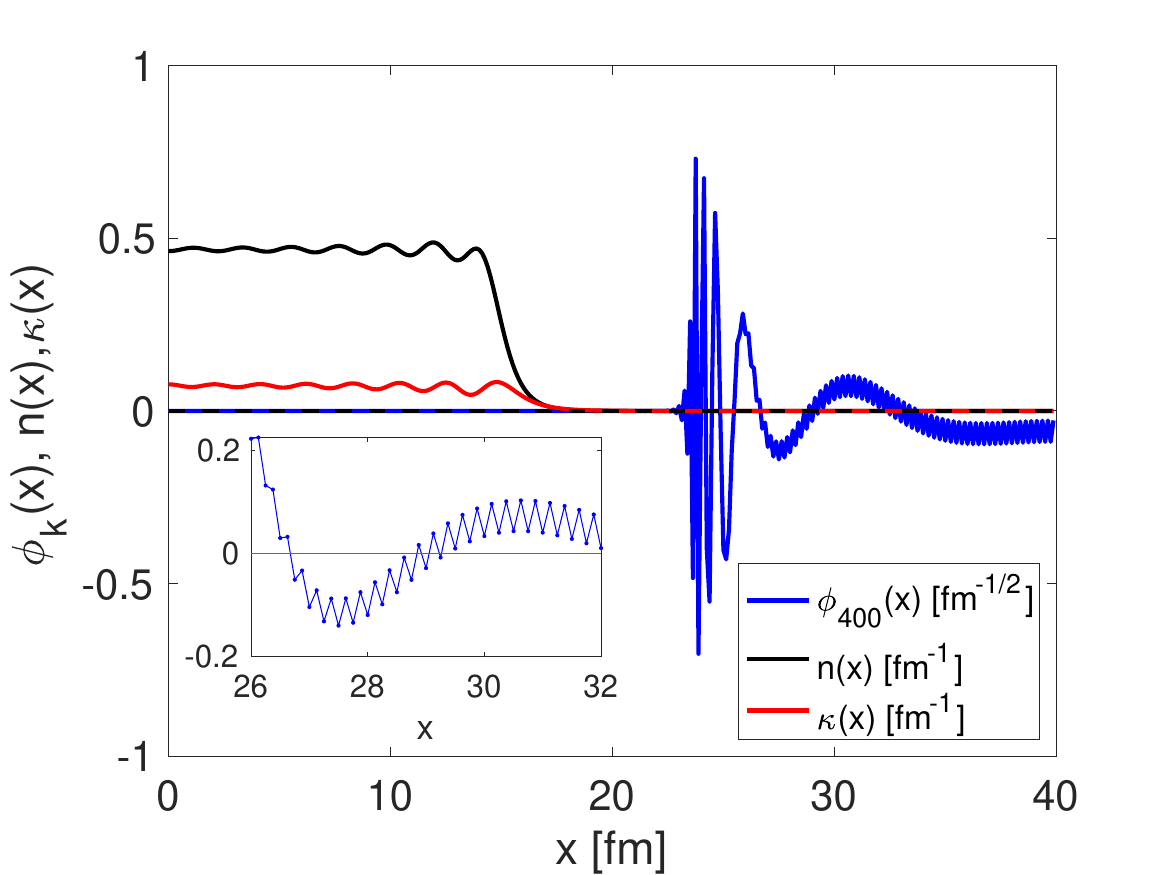}
\caption{ \label{fig:phi_300}   
The same as Fig.~\ref{fig:phi_12} for $\phi_{400}(x)$ and $n_{400}=2.6E-13$, 
an ``exterior'' canonical wave-function localized outside the system, with  $k$ beyond the UV-knee.
These type of wave functions clearly cannot be solutions of a typical 
Schr{\" o}dinger equation with a local potential. The inset shows that the high frequency 
spatial oscillations have a wave-length $2dx$, determined by the momentum cutoff $\Lambda$. In the limit $\Lambda \rightarrow \infty$ 
these canonical wave functions have no spatial derivatives, as they oscillate from lattice point-to-lattice point.}
\end{figure} 

The canonical occupation probabilities $n_k$ shown in Figs.~\ref{fig:number} and ~\ref{fig:nk} have a conspicuous behavior not
discussed previously in literature. For smaller lattice constant $dx$ the maximum momentum cutoff 
$\Lambda={\hbar\pi}/{dx}$  is large and the spectrum of $n_k$ extends to high  energies. The 
profile of $n_k$ has two obvious ``knees,'' one close to the Fermi level, the infrared IR-knee,  
where a transition from the BCS-like behavior of $n_k$ to a power-like behavior occurs for $k\approx 20$ in Fig.~\ref{fig:nk},
and a second one at a high energy, the ultraviolet UV-knee.   
The  canonical wave functions $\phi_k(x)$ have the expected spatial behavior as long as their support is commensurate with the support of
the number density matrix $n(x,y)$ as discussed above, see Eq.~\eqref{eq:nnn1}, the text below, 
and Figs.~\ref{fig:phi_12} and \ref{fig:phi_44} in the case of $dx=0.125$ fm.
However, as soon as the support of the canonical wave functions $\phi_k(x)$ is 
essentially outside the support of the density matrix $n(x,y)$, see Fig.~\ref{fig:phi_300}, 
for which  the index $k$ is on the right of the UV-knee in Fig.~\ref{fig:nk}, the corresponding $n_k$ decay significantly faster with $k$. 
Both the profiles and the numerical values of $n_k$ for these canonical states can be obtained with greater accuracy 
using increased precision, see Fig.~\ref{fig:nk}. These canonical
occupation probabilities do not identically vanish 
simply due to obvious quantum localization effects, but they are increasingly smaller 
with increasing resolution and decreasing lattice constant $dx$. In the limit $dx\rightarrow 0$ 
the UV-knee $\rightarrow \infty$ and at the same time the number of canonical states localized outside the system also tends to infinity. 
These non-localized canonical states however are irrelevant in describing the physical properties of the system.

\subsection{Impact of long momentum tails in 3D}

Below the IR-knee in Figs.~\ref{fig:nk}  and \ref{fig:occup_fission1} the canonical occupation probabilities have the 
expected BCS behavior~\cite{Bardeen:1957}. 
It is clear however that in between the IR-knee and UV-knee there is a region where the canonical occupation probabilities 
have a power law behavior. Such a behavior, due to the short-range character of the nuclear forces, was predicted in 1980 by \textcite{Sartor:1980} 
and recently put clearly in evidence experimentally in nuclei by \textcite{Hen:2014,Hen:2017,Cruz-Torres:2021}. 
Shina Tan~\cite{Tan:2008a,Tan:2008b,Tan:2008c} has proven analytically the emergence of this behavior for 
fermions interacting with a zero-range interaction in 3D. Nuclear pairing is typically simulated in theory with a $\delta$-potential, which naturally leads to a 
local pairing field $\Delta(\xi)$~\cite{Bulgac:1980}, similar to the case discussed here.  \textcite{Tan:2008a,Tan:2008b,Tan:2008c} has shown that  in the case of a zero-range interaction
asymptotically $n_k\propto C/k^4$ for {\bf any} many-body state. 
This power law behavior of the number density is directly related to the 
divergence of the anomalous density matrix Eq.~\eqref{eq:number_a}. 
In the case of a 3-dimensional system it was shown in Ref.~\cite{Bulgac:1980} that the anomalous density matrix
$\kappa(\xi,\zeta) \propto {1}/{ |{\bm r}-{\bm r}'| }$ when $|{\bm r}-{\bm r}'|\rightarrow 0$, where ${\bm r}$ and ${\bm r}'$ 
are the spatial components of $\xi$ and $\zeta$ respectively. 

It is also important to appreciate that the presence of the 
long momentum tail $n_k=C/k^4$ implies  
\begin{align} 
\int_\Lambda^\infty  dk k^2  n_k =\frac{C}{\Lambda},
\end{align} 
and therefore the particle number converges rather slowly as a function of the upper momentum cutoff $\Lambda$. 
The actual particle number can be reproduced in mean field calculations simply by adjusting the chemical potential, 
thus introducing errors  in the actual value of the chemical potential (a correction which might be small in practice). 
On the other hand the total kinetic energy of a system
\begin{align} 
\int d^3k \frac{\hbar^2k^2}{2m} n_k
\end{align} 
obviously diverges if $\Lambda \rightarrow \infty$. Consequently, the correct evaluation of the total kinetic energy, 
and as a result  the evaluation of the total energy of a many-body system
becomes a rather subtle problem, which in modern theoretical nuclear physics is resolved
using the methods of effective field theory (EFT), where infinities are handled with ``kid-gloves.'' 
While within EFT one can define the total energy of the system,
the separate definitions of either the kinetic, interaction, and even separate parts of the interaction energies become meaningless.

A closer analysis of Fig.~\ref{fig:phi_44} clearly show that some canonical wave functions $\phi_k(x)$ oscillate much faster than the density $n(x,x)$. 
The oscillation of the number density $n(x,x)$ is due to confinement in a finite box, a finite Fermi momentum, and a relatively well defined 
Fermi momentum $k_F$, and is a behavior known for decades for all finite Fermi systems.
Our initial ``naive'' estimate of the maximum expected number of relevant canonical wave functions, see Eq.\eqref{eq:Nmax}, is an underestimate, since
the maximum momentum cutoff ${\hbar\pi}/{dx} >  \sqrt{2m|U|}$. The coupling of the qpwfs components ${\textrm v}_k(\xi)$ to the 
continuum states, facilitated by $\Delta(\xi)$, leads to spatial oscillations with any wave-vector up to the maximum allowed value $\hbar \pi/dx$. 
In the limit $dx\rightarrow 0$ 
the cardinality of the set of canonical wave functions $\phi_k(\xi)$ is either $\aleph_0$ for a finite system in a finite volume
or $\mathfrak{c}$ for an isolated finite system  in vacuum.  Therefore, one should use for the best estimate of the number $N_\text{max}$, 
the cutoff momentum $p_\text{max}={\hbar\pi}/{dx}$, and 
from the condition of accommodating a standing wave in our ``square well'' potential, with $2R\approx 14.9$ fm in our numerical example, one obtains 
the approximate position of the UV-knee at  $k_\text{max} ={2R}/{dx} +{\cal O}\left({a}/{R}\right ) \approx 240$ for $dx = 0.125$ fm
(as $k$ counts the number of half wave-lengths inside the potential well), 
in perfect agreement with our numerical identification of the UV-knee in Fig.~\ref{fig:nk}.  
When coupling a bound state through the pairing field $\Delta$ with the continuum, the strength of the bound state is spread
over a large energy range with very long tails, with a Lorentzian shape of the spectral distribution~\cite{Bulgac:1980}.  Moreover, in 
time-dependent phenomena, even in the absence of a true pairing condensate (when the long range order is lost) and at 
high excitation energies (with corresponding temperatures well above the pairing phase transition $T_c$) the remnant 
pairing field leads to many single-particle transitions and the quantum Boltzmann one-body entropy increases considerably~\cite{Bulgac:2022}.

With this in mind one can now provide a better estimate of the size of the canonical basis set for a 3-dimensional system in a finite 
simulation box with sides of length $L_x=N_xdx,\, L_y=N_ydy,\, L_z=N_zdz, \, (dx=dy=dz$), ignoring spin and isospin degrees of freedom,
\begin{align} 
N_\text{max} = \frac{4\pi}{3}\left( \frac{\hbar \pi}{dx} \right )^3 \frac{4\pi}{3}r_0^3A \frac{1}{(2\pi \hbar)^3} = 2.2 A\left (\frac{r_0}{dx}\right )^3.
\label{eq:NewNmax}
\end{align}
At the same time the total number of single-particle quantum states in such a box is
\begin{align}
N_\text{spwfs} =  \frac{L_xL_yL_z}{(2\pi \hbar)^3}\left( \frac{2\pi\hbar }{dx}\right )^3= N_xN_yN_z,
\end{align} 
which is typically significantly larger. For example for a typical simulation box for a heavy nucleus with volume $30^3$ fm$^3$ and $dx = 1$ fm 
the total number of qpwfs  (here ignoring spin and isospin degrees of freedom) 
is  $N_\text{spwfs}= 27,000 \gg N_\text{max}\approx 3.8 A<1,000$. 
This estimate is accurate only for some quantities, such as particle number and total energy, see section~\ref{section:IV}.

The classification of the wave functions as ``interior,'' as in Figs.~\ref{fig:phi_12} and \ref{fig:phi_44}, and ``exterior,'' 
as in Fig.~\ref{fig:phi_300}, depends on the momentum cutoff $\Lambda$, particularly when discussing the 
entropy of a quantum state, see section~\ref{section:IV}, and less so when evaluating the total energy of a system.  
Various sizes of sets of the canonical wave functions, 
with $k$ smaller than the IR-knee, have been considered in the evaluation of the binding energies of   
nuclei~\cite{Stoitsov:1993,Reinhard:1999,Dobaczewski:1996,Tajima:2004,
Tichai:2019,Robin:2021,Hoppe:2021,Fasano:2022,Chen:2022,Hu:2022,Hagen:2022,Kortelainen:2022,Tichai:2022} 
and they missed the long momentum tails discussed in this work and their relevance. Moreover, for unclear reasons, when diagonalizing the one-body density matrix
these authors obtained negative canonical occupation probabilities, while it is obvious that 
the one-body density matrix is a non-negative definite Hermitian operator.

\section{The orbital entanglement entropy for a system of  identical particles}\label{section:IV}

There exists a number of approaches in the literature for the definition of the orbital entanglement entropy in the case of indistinguishable 
particles~\cite{Zanardi:2002,Shi:2003,Lo-Franco:2016,Benatti:2020,Johann:2021}, which typically depend on the single-particle basis used. 
One can often find similar either explicit or implicit statements, see e.g. Ref.~\cite{Robin:2021,Tichai:2022}, 
that orbital entanglement entropy is basis dependent.  
This amounts to the statement  that the orbital entanglement entropy 
corresponding to a many-body wave function  depends on whether one uses 
harmonic oscillator wave functions or plane waves for example. Since different choices would lead to different values of the orbital entanglement 
entropy  it is not clear what would be the use of such a definition, 
as  clearly it will not represent some intrinsic property of the many-body system. 
However, this dilemma is easily resolved if one realizes that 
for an arbitrary many-body wave function, there is a unique definition of the single-particle orbitals, either the natural orbitals introduced by 
\textcite{Lowdin:1956a,Lowdin:1956} or the mathematically identically definition used to introduce canonical single-particle  wave functions 
in the case of superfluid systems. These sets and the properties of the single-particle wave functions are basis independent and are uniquely
defined by the many-body wave-function. Together with the well-established mathematical proof due to \textcite{Hohenberg:1964}, a proof which withstood 
the test of time,  that there is a one-to-one correspondence between the many-body wave function and the one-body density, and 
thus with the one-body density matrix,  makes it obvious that the set of canonical wave functions or natural orbitals have an intrinsic value. 

The motivation for introducing the definition used in Refs.~\cite{Zanardi:2002,Shi:2003} for the orbital entanglement entropy, 
was motivated by quantum computing applications, in which case one deals 
with well defined single-particle orbitals corresponding to the specific physical realization of qubits, 
which are not necessarily the same as the needs of QIS. The information encoded in a many-body wave function is not
identical with the information encoded in a specific representation of the same wave function in a chosen physical 
realization of a quantum computer. This is equivalent to the statement that the representation of many-body wave function 
in terms of Slater determinants formed from single-particle wave functions is basis dependent.

The question of whether an arbitrary many-body wave function is representable either by the corresponding one-body
 density~\cite{Hohenberg:1964,Dreizler:1990lr,Gross:2006,Gross:2012} or by its one-body density matrix~\cite{Coleman:1963,Coleman:1963a} 
has been discussed and resolved a long time ago. The definition of the 
one-body density matrix Eq.~\eqref{eq:ntr1} is valid for either a stationary or time dependent many-body wave function $\Phi(t)$, 
with either well defined particle number or not, and its representation through its eigenstates, here for the more general case of a time
dependent system, is invariant with respect to an arbitrary (time-dependent) unitary transformation ${\cal U}(t)$,
\begin{align}
&n(\xi,\zeta,t) = \langle \Phi(t) |\psi^\dagger(\zeta) \psi(\xi) | \Phi(t) \rangle,\label{eq:nt}\\
&\int d\zeta \, n(\xi,\zeta,t) \phi_k(\zeta,t) = n_k(t) \phi_k(\xi,t), \label{eq:ES}\\
&n(\xi,\zeta,t) = 
 \sumint_k v_k^*(\xi,t)v_k(\zeta,t),\\
&v_k^*(\xi,t) = \sqrt{n_k(t)}\phi_k(\xi,t),\\
&\overline{v}_k^*(\xi,t) = \sumint_l {\cal U}_{kl}(t)v_l^*(\xi,t).
\end{align}
In particular, the time-dependent Hartree-Fock-Bogoliubov (HFB) equations are invariant as well with respect to such unitary transformations. 
In other words if at some time $t$  the set of quasi-particle wave functions $v_k(\xi,t), u_k(\xi,t)$, see the appendix,  happens to be the canonical set, 
in general as time evolves the quasi-particle wave functions will not remain canonical~\cite{Bulgac:2019e}. The
time-dependent number and anomalous densities are invariant with respect to 
such time-dependent unitary transformations~\cite{Bulgac:2021c}, while the canonical occupation probabilities 
are always uniquely defined. 
Therefore, unlike the case of the stationary HFB equations, 
one cannot uniquely relate the quasi-particle wave functions with the eigenvalues of the corresponding HFB equations. 
This is major difference with the Hartree-Fock problem, in both its time-dependent and stationary formulation. 
In the case of a time-dependent  many-body wave function one should introduce at each time the instantaneous occupation 
probabilities $n_k(t)$, see Eqs.~(\ref{eq:nt}, \ref{eq:ES}),  in order to have a unique definition  
of the time-dependent orbital entanglement entropy $S(t)$
\begin{align} 
S(t) = &- g\sumint_k n_k(t)\ln n_k(t)  \nonumber\\
&- g\sumint_k  [1-n_k(t)]\ln[1-n_k(t)].\label{eq:entt}
\end{align}

The usefulness of the orbital entanglement entropy becomes clear particularly in the case of non-equilibrium processes~\cite{Milburn:1997,
Chuchem:2010,Cohen:2016,Abanin:2019,Sinha:2020,Del-Maestro:2021,Del-Maestro:2022,Thamm:2022}. 
In the limit of a dilute and weakly interacting system the orbital entanglement entropy $S(t)$,
becomes a very good approximation of the time-dependent 
non-equilibrium thermodynamic entropy of a many-body system~\cite{Nordheim:1928,Uehling:1933}, similar to the case of the classical 
Boltzmann equation, see discussion in Ref.~\cite{Bulgac:2022}.  

Using the definition of the orbital entanglement entropy through canonical or natural orbital occupation probabilities 
the HF many-body wave function always has a vanishing orbital entanglement entropy. 
The canonical or natural orbital occupation probabilities 
offer a natural, unique, and simple way to characterize the entanglement properties of systems of indistinguishable 
particles. As has been mathematically proven~\cite{Coleman:1963,Coleman:1963a,Davidson:1972}, using natural orbitals,
an arbitrary many-body wave function has a well defined and unique Schmidt 
decomposition,  which thus allows a unique way to introduce the orbital entanglement entropy, 
irrespective of the single-particle basis used. The canonical or natural orbital occupation probabilities, 
which are obtained after the Schmidt decomposition of a many-body wave function, 
in order to construct the so called entropy spectrum~\cite{Li:2008}, 
can and do play a great role in characterizing topological phases of matter.

\section{Complexity of the many-body wave-function in the case of a non-equilibrium process}\label{section:V}

Both the quantum Boltzmann one-body and Shannon entropies can be evaluated only after the evaluation of the canonical occupation probabilities, 
see Eqs.~\eqref{eq:ent} and (\ref{eq:Shannon}). Both these entropies reach their minimal values only in the 
case of pure Slater determinants. Only in the presence of interparticle interactions these entropies increase in value
and can provide a measure of the complexity of the many-body wave-function. As far as we are aware 
there exist no studies of how the complexity, or the degree of single-particle spreading over the entire spectrum, depends on real time
in the case of a non-equilibrium process, particularly for a system with a high degree of complexity.  
Nuclear fission is a particularly interesting case and it provides an unexpected insight into how
the many-body wave-function evolves in time within the DFT framework from a state near the outer saddle point until 
after the two fission fragments are fully separated. 

\begin{figure}
\includegraphics[width=1.05\columnwidth]{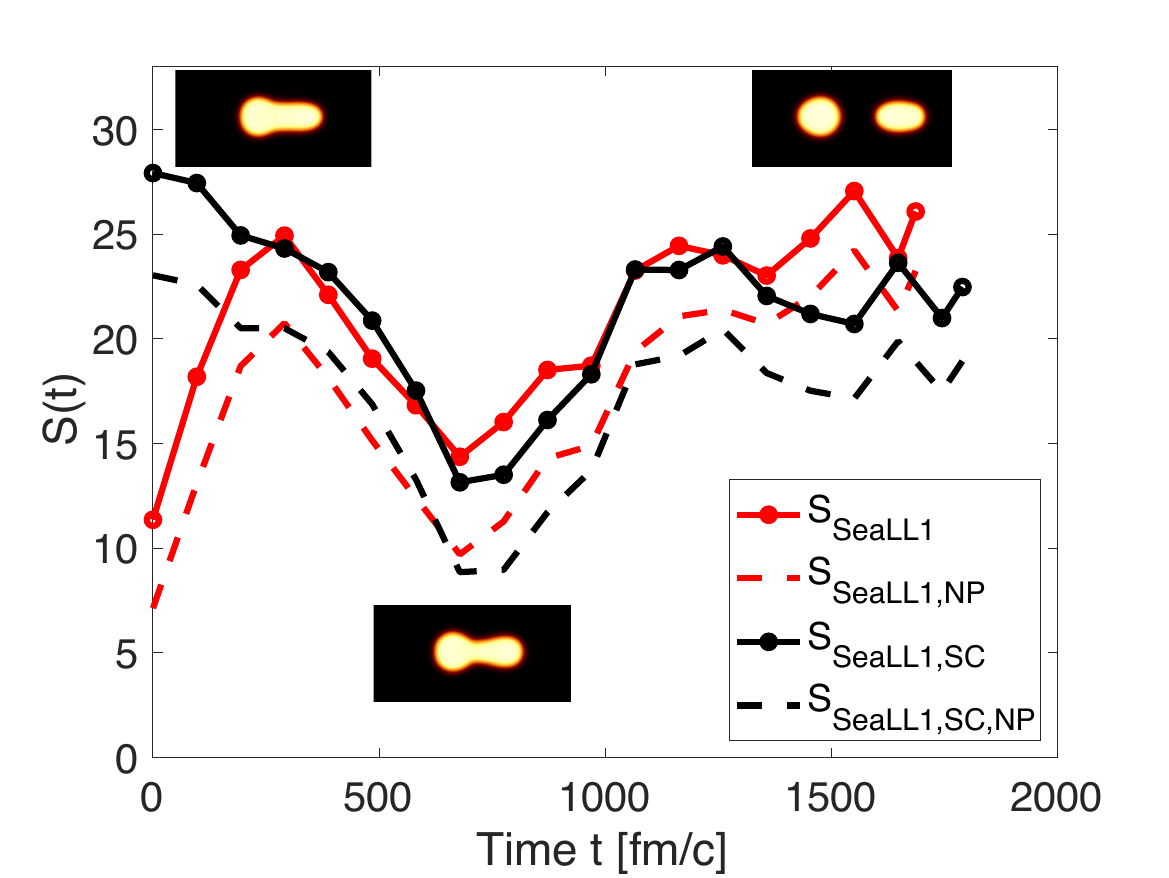}
\caption{ \label{fig:entropy}  
The time-dependence of the entropy $S(t)$ evaluated in the case of the  induced fission of $^{ 235}$U(n,f)
with a low energy neutron as a function of time from the vicinity of the outer saddle point until the two fission fragments are fully separated.
The solid curves correspond are entropies evaluated without particle projection of the total many-body wave-function, 
while  the dashed curves are obtained after particle projection was performed before the canonical occupation probabilities were evaluated.
The difference between the black and red curves is due to the difference between the initial states. In the case of the red curves we used
initial densities obtained by solving the constrained self-consistent HFB equations using the code HFBTHO~\cite{Navarro:2017,Marevic:2022}
using a relatively small set of transformed harmonic oscillator basis states. 
In the case of the black curves we obtained constrained 
self-consistent solutions directly on a 3D spatial lattice , which corresponds to a much larger single-particle space and a high momentum 
cutoff $\Lambda=\hbar \pi/ l \approx 600$ MeV/c, where $l=1$ fm is the spatial lattice constant and the dimension of the HFB matrix in this case is 216,000.
Even though there are differences between the values of the entropies evaluated before and after a total particle 
projection was performed, the qualitative behavior of the quantum Boltzmann one-body entropy is by and large 
the same.  The nuclear shapes obtained in TDSLDA during the time evolution are shown at 0, 675, and 1650 fm/c.
}
\end{figure}

Nuclear fission is a typical example of a non-equilibrium quantum process and 
one would expect that the entropy would monotonically increase in time~\cite{Bulgac:2022}. The actual situation however is more complex.  
We remind the reader that the entropy $S$ defined in Eq.~\eqref{eq:ent} is an entanglement entropy~\cite{Srednicki:1993,Klich:2006,Boguslawski:2014,Robin:2021}, 
which does not vanish even in the ground state of an interacting  many fermion system, unlike the thermodynamic entropy. In Fig.~\ref{fig:entropy} we display 
the time-dependence of the entropy $S(t)$ evaluated in several different manners, for initial conditions obtained in different methods and 
with and without particle number projection within DFT extended to superfluid 
systems~\cite{Bulgac:2002,Bulgac:2002a,Bulgac:2007,Bulgac:2011,Bulgac:2013a,Bulgac:2019}. 
We evaluate here the neutron and proton canonical occupation probabilities as a function of time, for both unprojected and projected total proton and neutron numbers,
following the techniques described in Refs.~\cite{Bulgac:2019,Bulgac:2021c} and illustrate the time evolution of 
the orbital entanglement  entropy in the case of $^{235}$U(n,f) induced fission with a low energy neutron, 
described with the nuclear energy density functional 
SeaLL1~\cite{Bulgac:2018} and using the code LISE~\cite{Shi:2020}. This extension of DFT to superfluid fermion system, in the spirit of the  local density Kohn-Sham framework,
is called the Time-Dependent Superfluid Local Density Approximation (TDSLDA).

\begin{figure}
\includegraphics[width=0.88\columnwidth]{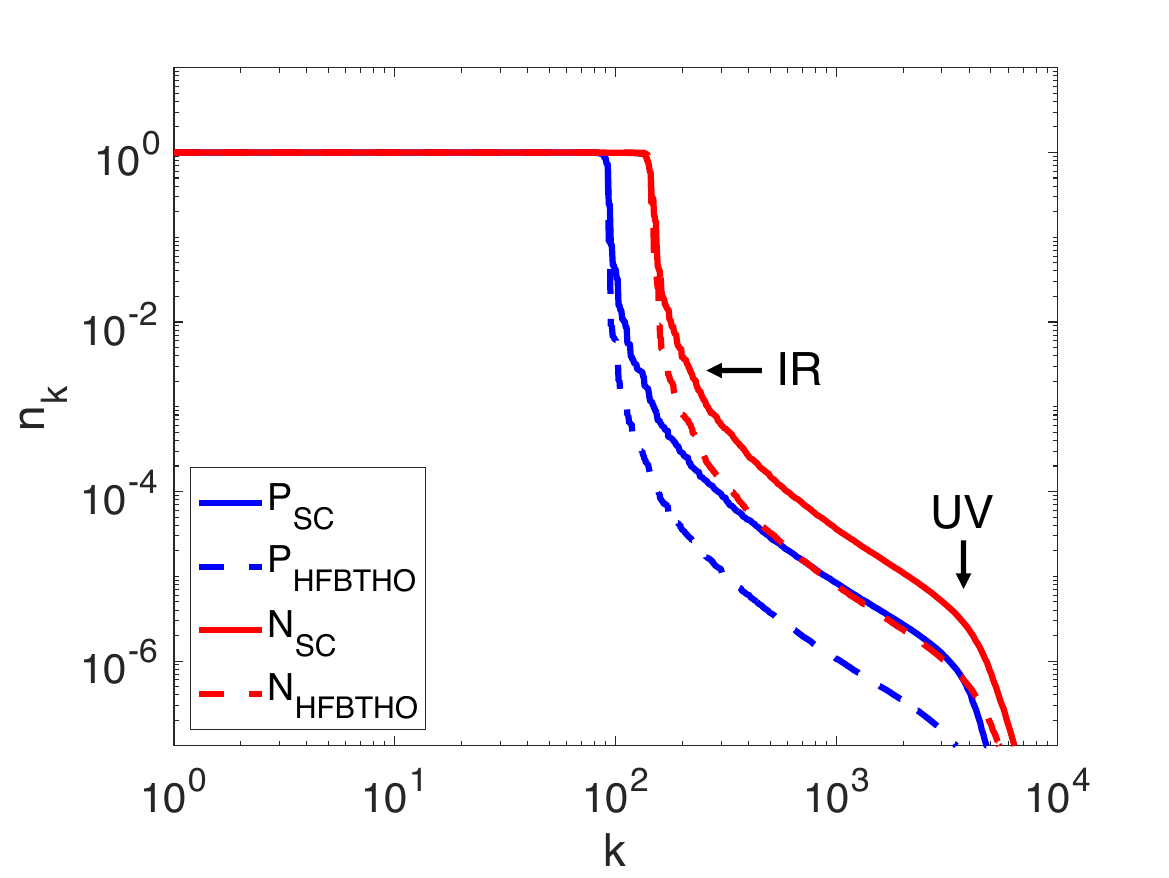}
\caption{ \label{fig:occup_fission1}  
The canonical occupation numbers at $t=0$ obtained from HFBTHO and self-consistent SLDA solution on the 3D spatial lattice
for protons and neutrons respectively in the case of $^{236}$U induced fission. Here $n_k$ are ordered in decreasing order. 
The canonical occupation probabilities up to a constant define the entanglement spectrum
 $-\ln n_k$~\cite{Li:2008}.}
\end{figure}

\begin{figure}
\includegraphics[width=1.0\columnwidth]{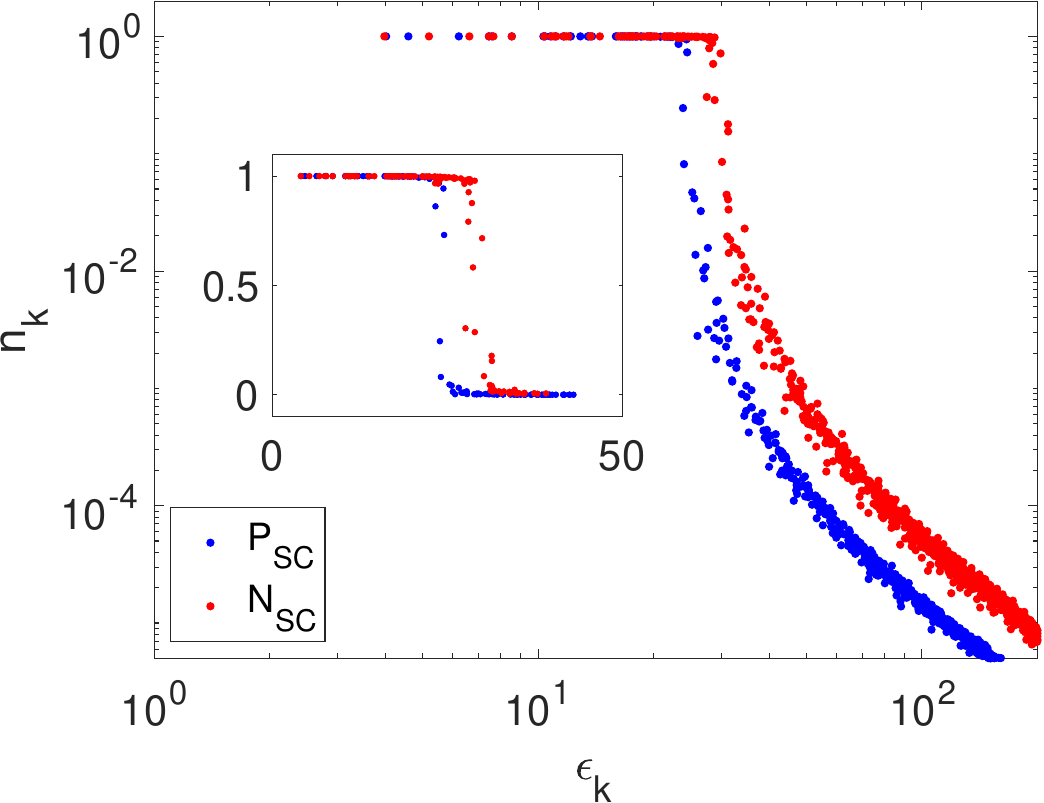}
\caption{ \label{fig:ekin}  
The canonical occupation probability $n_k$ as a function of 
$\epsilon_k$. 
In the inset we show that the canonical occupation probabilities $n_k$ around the Fermi level have the 
expected textbook behavior. Comparing Figs.~\ref{fig:occup_fission1} and \ref{fig:ekin} one sees that $\epsilon_k$ 
are only approximately monotonic functions of $k$, which, only for relatively large values of $\epsilon_k$, can 
be related to eigenstates in the approximately square well nuclear self consistent 
potentials  for different angular momenta~\cite{Brack:1997}.  }
\end{figure}

Beyond the UV-knee, for the canonical states localized mostly outside the system, the mean kinetic energies 
$\epsilon_k$ drop  in value and their contribution to the total kinetic energy is commensurate with what 
one expects from numerical discretization  errors ($dx\neq 0$)  of the continuum. As one can see from  Fig.~\ref{fig:ekin}, 
in  the region between the IR- and UV-knees, see Fig.~\ref{fig:occup_fission1}, the canonical occupation probabilities have
the approximate expected behavior $n_k\propto 1/\epsilon_k^2$, where 
\begin{align}
\epsilon_k=\left \langle \phi_k \left | -\frac{\hbar^2 {\bm \nabla}^2}{2m}\right | \phi_k\right \rangle .
\end{align} 
Since nuclear systems are to a large extent saturating systems, while the linear momentum ${\bm p}=-i\hbar{\bm \nabla}$ 
is not conserved, its absolute value is rather well defined and the single-particle wave functions can be well approximated 
in the semiclassical limit, the single-particle energies can be evaluated by quantization of classical orbits,  
and the shell structure of both spherical and deformed systems is reproduced with impressive accuracy~\cite{Brack:1997}. 
 
\begin{figure}
\includegraphics[width=1.0\columnwidth]{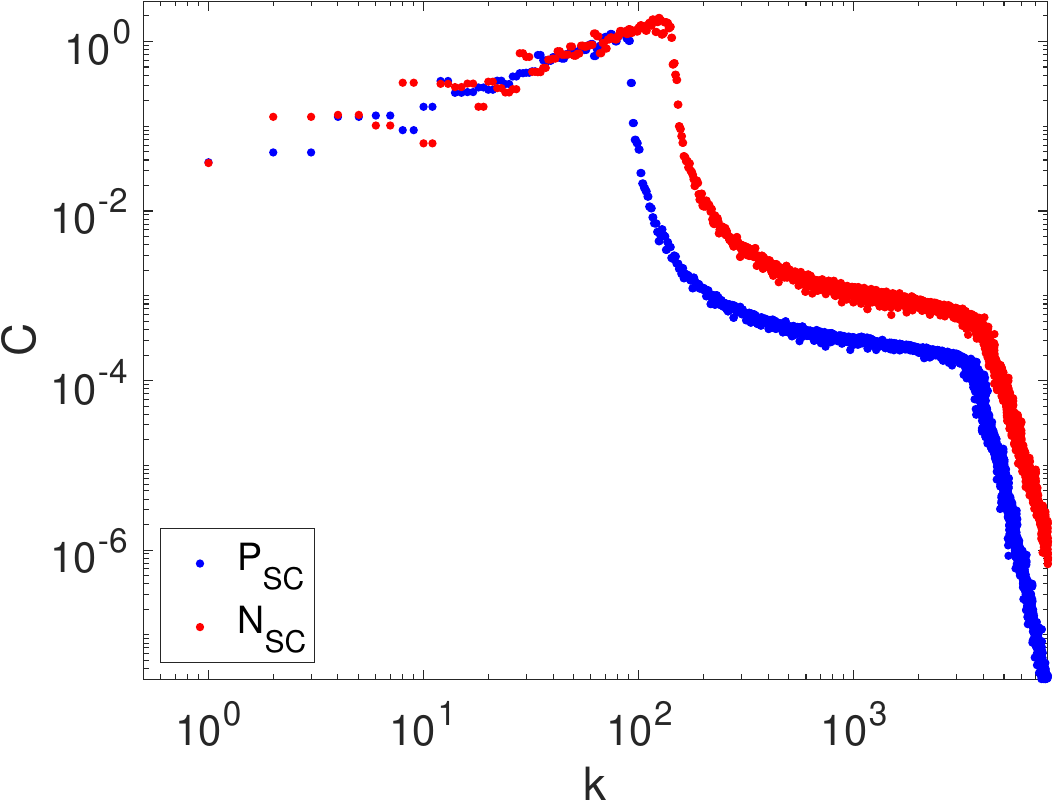}
\caption{ \label{fig:ekin1}  
The  quantity $C= (2m\epsilon_k/\hbar^2)^2n_k$, which in the regime $n_k\approx C/k^4$, between the IR- and UV-knees,
defines Tan's contact.   }
\end{figure}

\begin{figure}
\includegraphics[width=0.88\columnwidth]{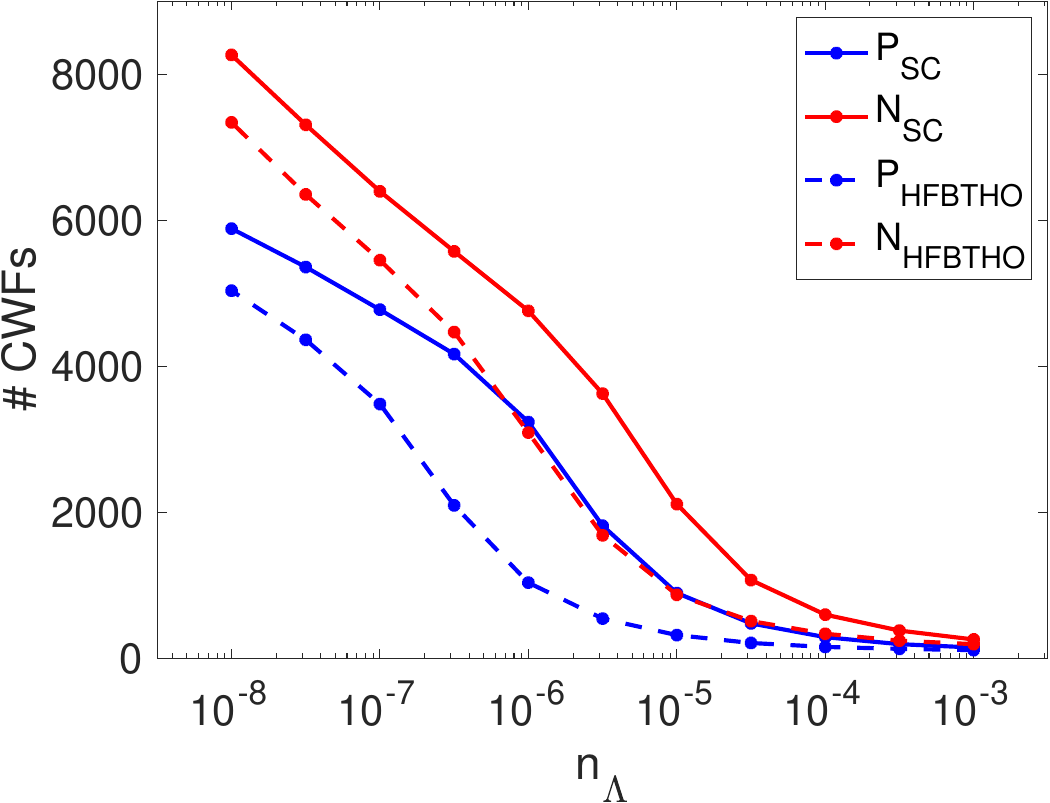}
\caption{ \label{fig:occup_fission2}  
 The number of canonical wave functions at $t=0$ as a function of the canonical occupation number cutoff 
$n_\Lambda$.  The dashed lines correspond to the HFBTHO set of self-consistent solutions, while the solid line 
correspond to the self-consistent solutions obtained on the 3D spatial lattice.   }
\end{figure}  

As we have proceeded in all our past TDSLDA simulations of nuclear fission~\cite{Bulgac:2016,Bulgac:2019c,Bulgac:2020,Bulgac:2021,Bulgac:2022b}, 
the initial state was determined using the HFBTHO code~\cite{Navarro:2017, Marevic:2022} which uses a small single-particle set of 
wave functions of size, which is quite sufficient to estimate the total energy of a nucleus. 
Since the TDSLDA simulations are performed on a 3D spatial lattice $N_xN_yN_z= 30^2\times60$, 
with a lattice constant $l=1$ fm, the size of the HFB matrix is much larger $4\times30^2\times60= 216,000$ for neutrons and protons respectively. 
 
\begin{figure} 
\includegraphics[width=0.88\columnwidth]{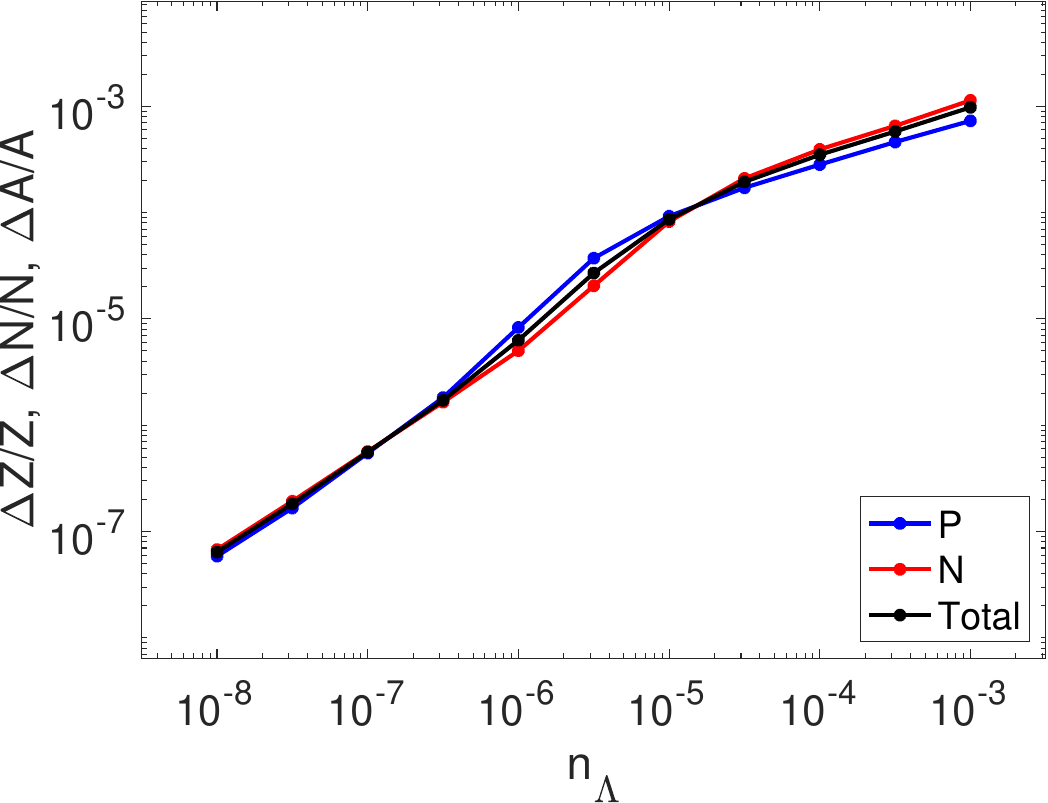}
\caption{ \label{fig:occup_fission4}  
 The accuracy of the particle number evaluated for the 
  self-consistent solutions at $t=0$ as a function of the canonical  occupation number cutoff.  }
\end{figure}

In Fig.~\ref{fig:occup_fission1} we show the canonical occupation numbers,  up to the UV-knee only,
obtained using the HFBTHO self-consistent densities and a set of self-consistent solutions on the 3D spatial lattice at $t=0$.
If one is interested in the total particle number the sum 
$N=\sum_k n_k$ converges with an accuracy 0.01 particles if summed up 
to $n_k\approx 10^{-5}$, see below the discussion of Fig.~\ref{fig:occup_fission4}, thus at most a few thousands canonical states 
(both spin-up and down) in the case of the self-consistent solution on the 3D  spatial lattice, 
a number almost an order of magnitude smaller than the size of the basis set $2\times30^2\times60=108,000$ (the factor 2 is for the spin).  

Within SLDA, or any treatment of pairing correlations with a local pairing field $\Delta(\xi)$, the theory requires  
regularization and renormalization~\cite{Bulgac:2002,Bulgac:2002a}. 
We have checked that indeed 
\begin{align}
(\epsilon_k)^2n_k \approx \left ( \frac{\hbar^2}{2m} \right)^2 C \quad \text{if} \quad k_\text{IR}<k<k_\text{UV}
\end{align}
 see Figs.~\ref{fig:nk}, \ref{fig:occup_fission1}, and \ref{fig:ekin1}, confirming the 
theoretical prediction of Refs.~\cite{Sartor:1980,Tan:2008a,Tan:2008b,Tan:2008c}. In the case of pure finite-range nucleon interactions, 
with no zero-range components, there is an upper momentum cutoff controlled by the interaction range. 
When treating nuclear systems as composed of proton and neutrons the typical momentum cutoff is $\Lambda \approx$ 600 MeV/c, 
which is related to the QCD chiral symmetry breaking scale $\Lambda_{\chi}$ controlled by the nucleon size, as it makes no sense to consider 
the interaction between two nucleons when their quark clouds strongly overlap.  Fig.~\ref{fig:ekin1} demonstrates that between the IR- and UV-knees 
the canonical occupation probabilities approach asymptotically the expected behavior $n_k\propto 1/\epsilon_k^2$, even though our cutoff 
momentum $\Lambda = \hbar\pi/dx\approx 600$ MeV/c is not sufficiently high, as the momentum interval between the IR- and the UV-knees 
covers less than an order of magnitude. In this rather small momentum interval the behavior of the 
canonical occupation probability is closer to $n_k \propto 1/\epsilon_k^{2.3\ldots 2.5}$.

In the case of the HFBTHO solution the number of relevant canonical states is at most 1,000 or so, see  Fig.~\ref{fig:occup_fission2} for the number of 
canonical wave functions up to a given occupation number cutoff $n_\Lambda\approx 10^{-5}$. A particle projected
many-body wave function can now be expressed as a sum of Slater determinants, built from canonical states/natural orbitals. 
The number of these Slater determinants can be considered as an appropriate measure
of the  complexity of a many-body wave function
\begin{align} 
\frac{N_{CW\!Fs}!}{N!(N_{CW\!Fs}-N)!} \ll \frac{N_{sp}!}{N!(N_{sp}-N)!} 
\end{align}
which is exponentially smaller than the total number of possible Slater determinants (for either neutrons or protons) 
in the entire many-body Hilbert  space corresponding to $N_{sp}=2\times N_xN_yN_z=108,000$. 
In the case of a shell-model or CI calculation for example, the complexity, and likely the accuracy as well,  
of the many-body wave function thus cannot be judged by the dimension of the many-body Hilbert space, which depends 
on the type of single-particle wave functions used.

\begin{figure}
\includegraphics[width=0.88\columnwidth]{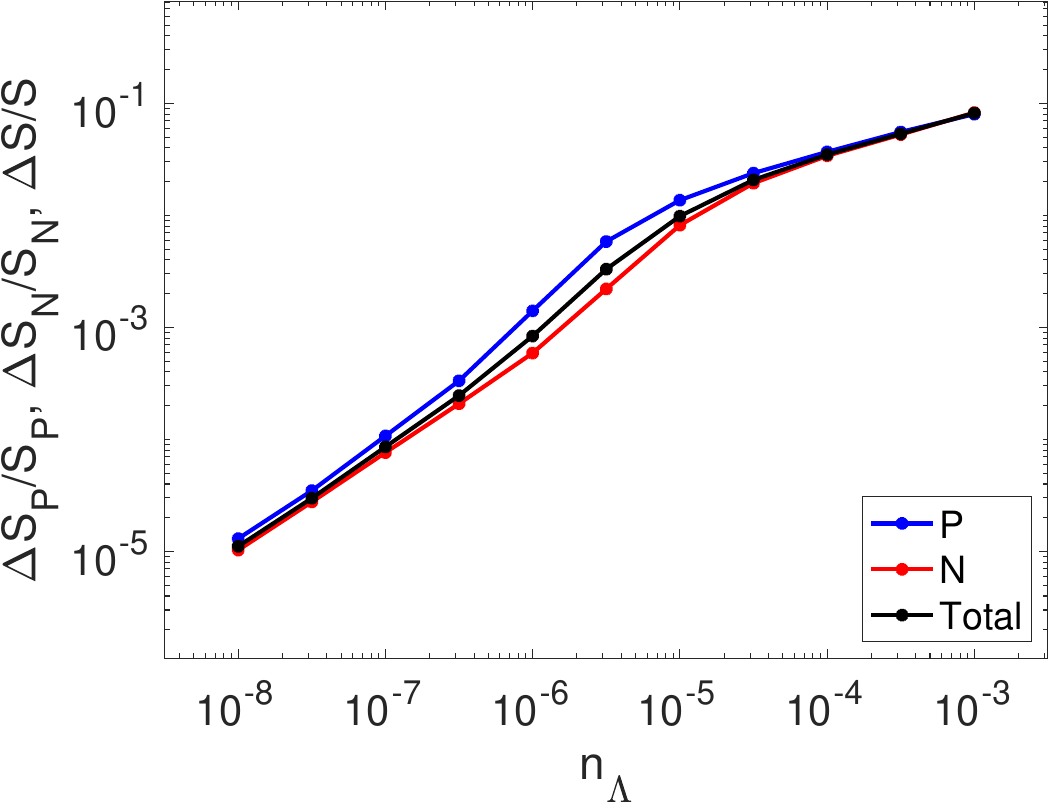}
\caption{ \label{fig:occup_fission3}  
 The accuracy of the entropy evaluated for the 
 self-consistent solutions at $t=0$ as a function of the canonical  occupation number cutoff.  }
\end{figure}

In the case of HFBTHO the chemical potentials 
can be tuned to fix the desired particle numbers, even if the size of the single-particle space is (artificially) small. 
The particle number and the total energy of the system converges faster as a function of the cutoff in $n_k$ when compared 
to the entanglement entropy $S(t)$, compare Figs.~\ref{fig:occup_fission4} and \ref{fig:occup_fission3}. 
The lesson is that one cannot judge the quality or the complexity of a wave-function by using a 
wave-function obtained by a variational estimate for a qualitatively different observable, 
see also the arguments presented by \textcite{Anderson:1984} and our discussion above.

In a full TDSLDA simulation of fission we have extracted the canonical occupation probabilities as a function of time, 
and for all times their qualitative behavior is very similar to that at $t=0$ as illustrated in Figs.~\ref{fig:occup_fission1}, 
and \ref{fig:ekin}, even though the pairing condensates are absent for times > 700 fm/c and the 
role of SRCs is always manifest, see also Refs.~\cite{Bulgac:2022,Bulgac:2022a}.
The entropies $S(t)$ evaluated using HFBTHO initial wave functions are shown with the red solid and dashed lines 
in Fig.~\ref{fig:entropy}, in the case of unprojected particle numbers and projected particle numbers respectively. 
The initial densities obtained with the code HFBTHO were placed on this spatial lattice and only the proton and neutron chemical 
potentials were slightly adjusted, in order to obtain the correct particle numbers $Z=92$ and $N=144$ respectively. 
Fully self-consistent initial wave functions obtained on the 3D spatial lattice $N_xN_yN_z=30\times30\times60$
were used to determine the canonical occupation probabilities for the entropies $S(t)$ shown with black solid and dashed lines
for unprojected particle numbers and projected particle numbers respectively. 
The difference between the initial canonical occupation probabilities obtained using the HFBTHO, in which only the chemical 
potentials were adjusted,  and fully self-consistent solutions obtained on the 3D  $N_xN_yN_z$ lattice are illustrated 
in Fig.~\ref{fig:occup_fission1}. In Fig.~\ref{fig:entropy} with dashed lines we present the corresponding entanglement
entropies evaluated after the proton and neutron particle projections were performed at each time.
 
\vspace{0.5cm}

After each 100 fm/c  the time-dependent neutron and proton 
density matrices were used to determine the instantaneous canonical occupation 
probabilities and evaluate the corresponding $S(t)$ shown in Fig.~\ref{fig:entropy}. 
In order to evaluate the entropy with an accuracy at the  $\approx$0.1\% level  in Fig.~\ref{fig:entropy} 
we needed to account for occupation probabilities with $n_k\ge 10^{-6}$.  
The entropies are larger in the case of particle unprojected wave functions, as such many-body 
wave functions have more complexity, 
since they contain  components with different particle numbers. At the same time, for
the evaluation of the total particle number, and implicitly of the number densities and total energy of the system, 
it is sufficient to include only states with $n_k\ge 10^{-4}$. In time-dependent simulations however, we have found that one cannot limit
the number of qpwfs included in the calculations, without severely affecting the outcome. A reformulation of the time-dependent 
DFT within a canonical wave-function basis set does not exist at this time. A major difficulty is that a set of initial canonical wave functions 
does not remain canonical under time evolution.

Our initial nuclear configuration corresponds to a nucleus slightly above the outer fission barrier, when the nucleus starts its evolution towards 
the scission configuration and the neck is formed. Scission occurs quite fast after a time $\approx $700 fm/c, after which the two fission fragments recede 
from each other, although their shapes still evolve and their equilibrium is attained at much later times~\cite{Bulgac:2019c,Bulgac:2020}. 
 The initial wave-function describes the ground state of a shape constrained 
nucleus, with thus technically zero thermodynamic many-body entropy. While the initial state was a ``bound'' state, the final non-stationary state lies in a continuum, 
where the density of many-body states is very large, even in the finite simulation box used in our numerical simulation. 

In the case of HFBTHO initial conditions the entropy of the nucleus increases up to a time $t\approx 300$ fm/c, 
matching the entropy of the system obtained with fully self-consistent initial conditions obtained on the spatial lattice. 
Once the initial HFBTHO conditions were placed on the spatial lattice the nucleus ``realizes'' that the full nuclear wave-function ``lives'' 
in a much larger space, the ``doors are opened wide''  and the system ``expands'' accordingly, 
until the HFBTHO initial conditions reach in time a complexity comparable to the complexity of the 
self-consistent many-body wave-function obtained on the 3D spatial lattice. 
We conclude that more accurate initial conditions are needed in the future studies, to eliminate the ``unphysical'' evolution 
caused by using HFBTHO initial conditions. 
We remind the reader that we routinely perform (TD)SLDA calculations
with a lattice constant $dx=1$ fm, which corresponds to a cutoff momentum 
$\Lambda=\hbar \pi/dx \approx 600$ MeV/c, which is the typical largest cutoff 
momentum used in chiral perturbation effective field theory studies of nucleon interactions. It is our hope that by including effective 
pn-pairing~\cite{Bulgac:2022} one can eventually simultaneously capture all long- and short-range correlations in a 
mean field like approach. 

The question however remains, why the actual entropy $S(t)$, the solid and dashed lines in Fig.~\ref{fig:entropy}, decreases until scission. The natural 
 explanation is that while the neck is forming,  ``communication'' between the emerging fission fragments is hindered and it completely stops after scission, thus
 the space in which the quasi-particle wave functions evolve becomes smaller than it was at the initial time.  When the neck starts emerging the free 
flow of nucleons from one side to the other of the elongated nucleus is increasingly inhibited until a time $\approx 700$ fm/c, when the 
neck attains a very small diameter and after that the two fission fragments start separating.  During this time period until $\approx$ 700 fm/c 
the spreading of the single-particle strength is suppressed.  After scission the two fission fragments emerge with significant 
 excitation energy, $\approx 15\ldots 20$ MeV each, but obviously not in thermodynamic equilibrium, as in particular their shapes evolve as well.  
These aspects are likely also connected with the widely studied problem of many-body localization, 
a topic of significant interest in predominantly 1D condensed matter systems~\cite{Abanin:2019}. When the 
hopping strength between sites is weaker than the amplitude of the disorder such 1D systems fail to thermalize. 
The forming of a neck between emerging fission fragments basically plays the same role as disorder in 1D 
condensed matter systems, it increasingly inhibits the nucleon jumps between these emerging fission fragments.
Unlike in the case of a system in contact with a thermostat, the entropy of an isolated many-body system, or in a more precise language, 
the complexity of its many-body wave-function does not necessarily always increase monotonically in time at intermediate times. A similar situation is
often used in demos in introductory physics classes, when compressed air is released from a container, it cools, 
and its entropy decreases.

\begin{figure}
\includegraphics[width=0.88\columnwidth]{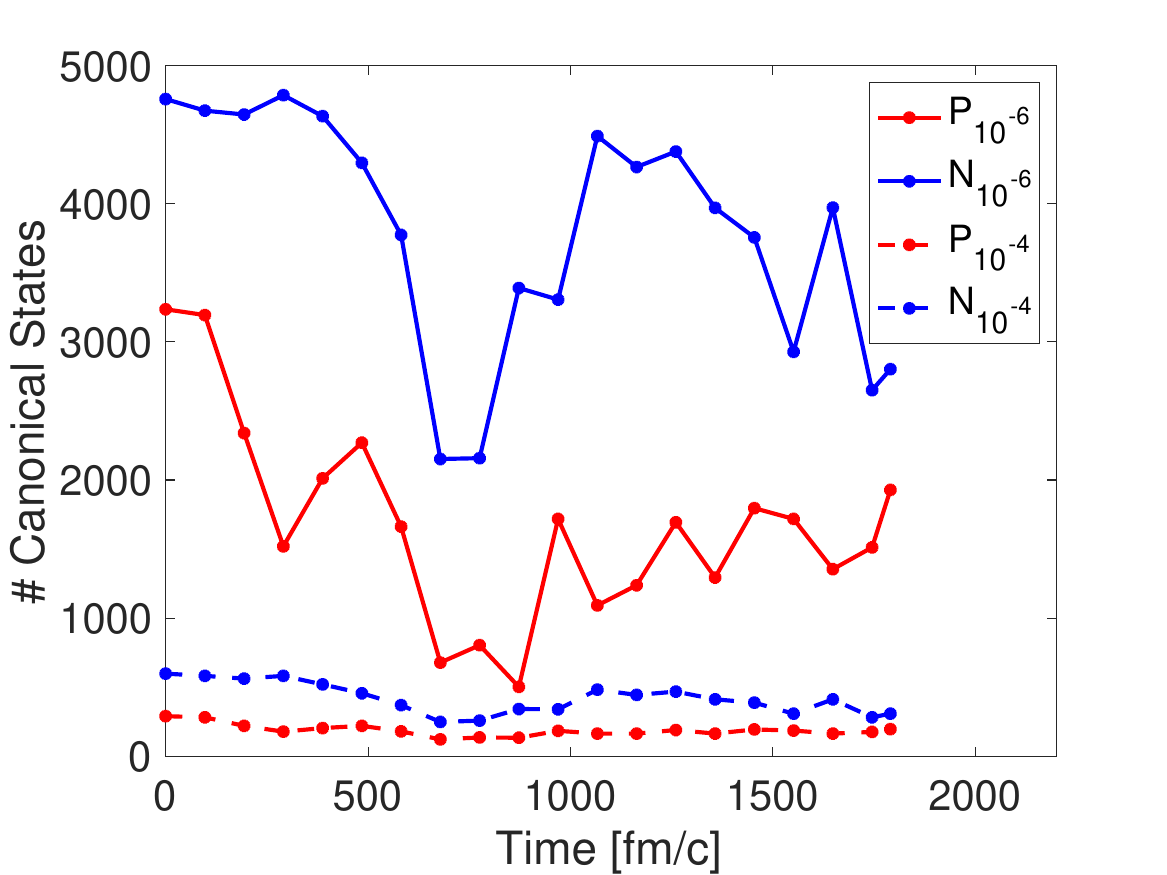}
\caption{ \label{fig:occup_fission5}  
The  number of canonical states vs time for cutoffs $n_k \le 10^{-4}$ and $10^{-6}$ respectively. 
These have to be compared with the size of the entire set of canonical wave functions 
$2N-xN_yN_z=108,000$ in this study. }
\end{figure}  

\begin{figure}
\includegraphics[width=0.88\columnwidth]{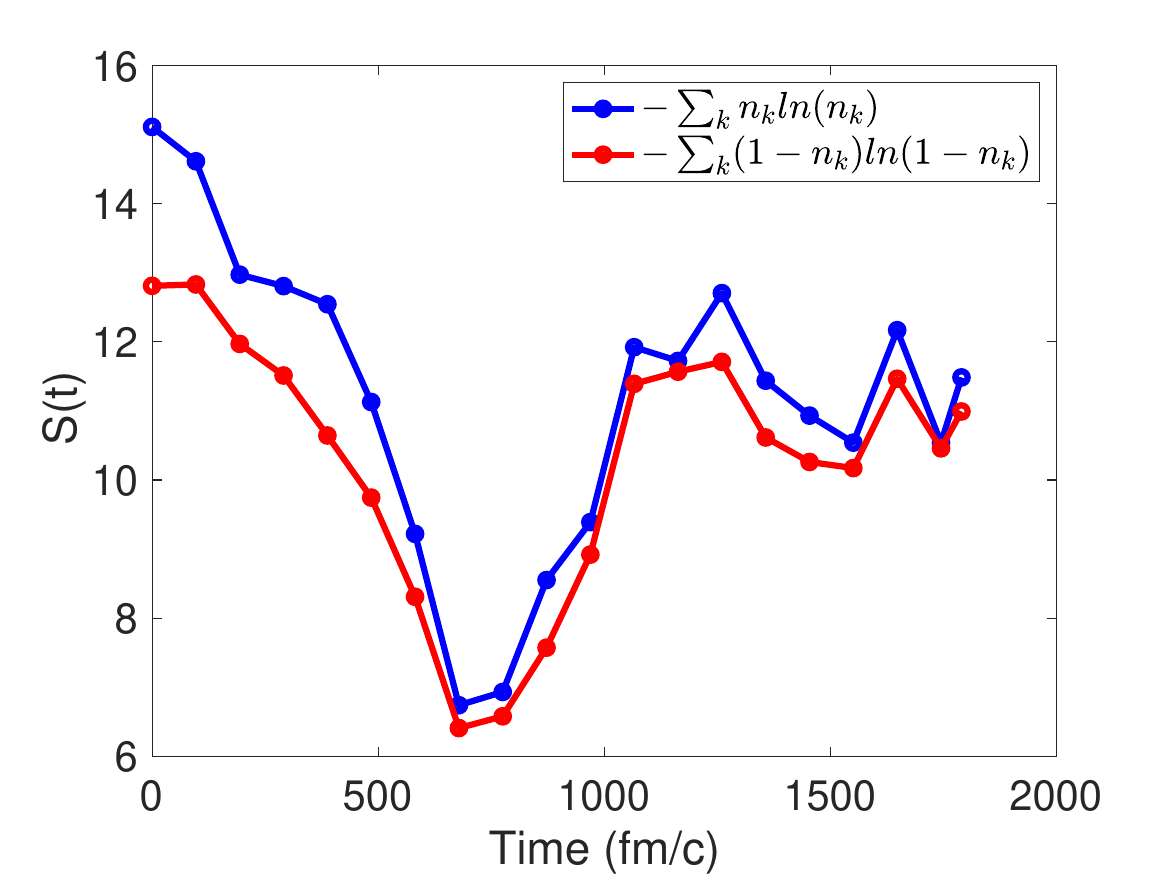}
\caption{ \label{fig:occup_fission6}  
The time-dependence of the two parts of the entanglement entropy 
$S(t) = S_1(t)+S_2(t)$, where $S_1(t)=-\sum_kn_k\ln n_k$ and $S_2(t)=-\sum_k(1-n_k)\ln(1-n_k)$. 
Up to an additive constant  and an overall normalization, $S_1(t)$ is equal to the Shannon entropy $\overline{S}$, see Eq.~\eqref{eq:Shannon}. }
\end{figure}  

It is instructive to determine the number of canonical states with occupation probabilities $n_k$ up to different cutoffs, see Fig.~\ref{fig:occup_fission5}. 
A cutoff in canonical occupation probabilities at the level $n_k\ge 10^{-5}$ can be sufficient for evaluating particle number with a relative 
accuracy ${\cal O}(10^{-4})$, see also Fig.~\ref{fig:occup_fission4}. 
However,  the evaluation of the entanglement entropy, which is a more accurate measure of the complexity of 
the many-body wave function, a cutoff in $n_k\ge10^{-6...-7}$ is needed, see Fig.~\ref{fig:occup_fission3}.   

It is also instructive to evaluate the two different contributions to the entanglement entropy $S(t)=S_1(t)+S_2(t)$, 
see Eq.~\eqref{eq:ent}.  As one can judge from Fig.~\ref{fig:occup_fission6} $S_1(t)$ and $S_2(t)$ have a very similar 
behavior and almost equal magnitudes. This is particularly relevant, since $S_1(t)$  is identical to the Shannon entropy $\overline{S}$, 
up to an additive constant and a multiplicative 
factor, see Eq.~\eqref{eq:Shannon}.  

One might hastily conclude that our conclusions are limited to the case of pairing correlations only. This conclusion would be wrong 
for several reasons. i) Even when the pairing correlations, not necessarily the pairing condensates, occur only in the $s$-wave between protons and neutrons
only, the corresponding pairing fields still describe the role of collisions.  At the semiclassical level such collisions are incorporated by  the collision integral 
in the extension of the Boltzmann equation due to Nordheim~\cite{Nordheim:1928} and Uehling-Uhlenbeck~\cite{Uehling:1933} 
at finite local temperatures, see also the discussion in Ref.~\cite{Bulgac:2022}. The results obtained so far in the study of numerous cold atom and nuclear 
systems amply demonstrate that when these systems are highly excited, well above the critical temperature $T_c$ for the onset of 
pairing condensates and their corresponding pairing correlations are absent, 
the high-momentum states are increasingly occupied as time goes on, see
Refs.~\cite{Bulgac:2011,Bulgac:2017,Bulgac:2019c,Bulgac:2020,Magierski:2022}. In the results 
illustrated in Fig.~\ref{fig:occup_fission6}, at times beyond 700 fm/c in the emerging fission fragments pairing condensates 
are absent and the temperature of the fission fragments are $T>T_c$.
ii) In a Kohn-Sham like extension of the TDDFT the role of finite-range
interactions can be incorporated only through local potentials, as non-local Fock like potentials 
will not be used in the foreseeable future in time-dependent simulations. For example, in Ref.~\cite{Hen:2017} it is shown how the role of the 
tensor interaction at high-momentum transfer can be emulated with a pure attractive $s$-wave interaction in the $pn$ triplet channel, 
which dominates the SRCs in nuclear systems.
iii) Note, that even in the Boltzmann equation and its semiclassical extension, the Boltzmann-Uehling-Uehlenbck equation, 
the collisions are always local in space and
describe the time-evolution of the one-body density matrix, similarly to TDDFT. Collisions in higher partial waves
will likely be incorporated by corresponding anomalous densities depending on various gradients 
of the quasi-particle wave functions, as is done currently also for the corresponding terms in the self-consistent Skyrme-like potentials.   
The $pn$-collisions, in particular play the role of collisions due to the tensor interactions,  and are still absent in current nuclear 
TDSLDA simulations, but they can and will soon be incorporated~\cite{Bulgac:2022}. 

In Fig.~\ref{fig:entropy} we show the entanglement entropy evaluated both before and after particle projections. 
For times larger than 700 fm/c the nuclear system is normal ($T>T_c)$, and these entropies increase with time 
as expected for a non-equilibrium process. 
The number of Slater determinants in the corresponding particle projected expansion of the 
highly-correlated many-body wave function is still increasing in time. This number can be estimated for either the proton 
or neutron systems to be order of $10^{120...140}$
using Eq.~\eqref{eq:orb} with the number of relevant single-particle orbitals extracted from Fig.~\ref{fig:occup_fission5}.
This estimate is orders of magnitude larger than any size ever attempted in CI calculations, which never take into account explicitly SRCs. 
As our own still unpublished results in larger spaces confirm, the number of canonical states/natural orbitals in a time-dependent situation
increase dramatically with the size of the simulation. As more single-particle states become available the number of accessible final states 
increases as well and the trajectory of the reaction changes accordingly, as one would naturally expect.
  
\section{Conclusions}\label{section:VI}

In order to study the complexity of a many-fermion wave function 
we have concentrated here on the properties of the canonical wave functions/natural orbitals basis set 
and the corresponding occupation probabilities. Unlike previous studies, which were limited only to the study
of canonical wave functions/natural orbitals only for the ground state of various systems and small 
basis sets of such states, we have extended our analysis to excited and in particular to many-fermion 
functions describing non-equilibrium processes and have shown that their properties are universal and 
general properties of the one-body density matrix, whether the system is static or time-dependents, see 
also~\cite{Bulgac:2022,Bulgac:2022a}. 
The canonical wave functions/natural orbitals fall basically into two distinct categories: {\it a)} The first group 
corresponds to wave functions with their support predominantly inside the nucleus. {\it b)} The second category 
corresponds to wave functions with their support outside the nucleus and in the limit of high spatial resolution 
($\Lambda \rightarrow \infty$) these wave functions have vanishing occupation probabilities and hence do not 
contribute to observables. We have dubbed these two categories as ``interior'' and ``exterior'' canonical wave functions. 
As far as we are aware, the existence and properties of these two sets of canonical wave functions/natural orbitals 
has not been discussed in literature. The ``interior'' wave functions fall into two subgroups: the first group 
corresponding to occupation probabilities described by well-known either Bardeen-Cooper-Schrieffer distribution 
in the presence of pairing correlations or the textbook distribution~\cite{Abrikosov:1963}
\begin{align} 
n(k) = \frac{1}{1+\exp[-\beta(\epsilon(k)-\mu)]},
\end{align} 
where $\beta=1/T$, and the second group which has a power-law behavior
\begin{align} 
n(k) = \frac{C}{k^4},
\end{align} 
due to SRCs~\cite{Tan:2008a,Tan:2008b,Tan:2008c,Zwerger:2011,Hen:2017,Bulgac:2022} in both cold atoms and nuclear systems, 
in particular due to the dominant role of the tensor interactions between protons and neutrons~~\cite{Quint:1986,Quint:1987,Sartor:1980,Frankfurt:1981,Frankfurt:1988, Benhar:1993,Ciofi-degli-Atti:1996,Pandharipande:1997,Piasetzky:2006,Sargsian:2005,Schiavilla:2007,Sargsian:2014,Hen:2017,
Zwerger:2011,Braaten:2011,Castin:2011,Anderson:2015,Porter:2017,Stewart:2010,Hen:2014,Hen:2017,Aumann:2021,Cruz-Torres:2021} 
in nuclear systems.
Moreover, while for an isolated quantum system in vacuum the  cardinality of the entire 
set of canonical wave function is $\mathfrak{c} = |\mathbb{R}^3|$, the cardinality of the ``interior'' subset of 
canonical wave functions is only $\aleph_0 = |\mathbb{Z}| = |\mathbb{N}|$. The spatial profile of these ``interior'' wave functions 
have the hallmark behavior of ``standing waves'' in finite potential well, with their ``spatial frequencies'' extending up to infinity
when $\Lambda\rightarrow \infty$. These ``interior'' states should not be confused with single-particle quantized 
states, which can be defined only in the case of static HF or HFB calculations, where the (generalized) density matrix commutes with the
(generalized) single-body Hamiltonian. 

In calculations performed in finite spatial boxes, the set of physically relevant ``interior'' canonical wave functions only 
has a significantly smaller size than the full set, which can be crucial in 
performing many-body simulations within such a reduced space, but ``physically complete'' set. 
The set of  $-\ln n_k$ is also known as the entanglement spectrum~\cite{Li:2008} and it is widely used in 
literature to characterize the properties of strongly interacting many-body systems~\cite{Chuchem:2010,Pal:2010,Bardarson:2012,Cohen:2016,Abanin:2019,Sinha:2020,Wimberger:2021, Liu:2022,Mueller:2022,Schneider:2022}.

The canonical basis set appears well suited for performing shell-model calculations~\cite{Johnson:2018}. However, 
it remains a challenge to reformulate TDDFT explicitly within this basis.  Approximate sets  of canonical states can 
be easily generated, for example, for nuclear problems, by solving the non-self consistent equations for the radial wave functions
\begin{align} 
&\begin{pmatrix} 
H  -\mu  & \Delta \\
\Delta   & -H^* +\mu
\end{pmatrix}
\begin{pmatrix}
{\textrm u}_{k} \\
{\textrm v}_{k}
\end{pmatrix}
= E_k
\begin{pmatrix}
{\textrm u}_{k}\\
{\textrm v}_{k}
\end{pmatrix},\\
&H=-\frac{\hbar^2{\bm \nabla}^2 }{2m} +V+V_{so},
\end{align}
where the central potential $V$ and the pairing field $\Delta$ have spherical symmetry, and $V_{so}$ is an appropriate 
single-particle spin-orbit  potential. The generation of sufficiently 
large sets of canonical wave functions, or natural orbitals, with exact quantum numbers $njlm$, 
in the case of spherical symmetry, is numerically cheap and the set can be adapted 
for the problems studied in Refs.~\cite{Stoitsov:1993,Reinhard:1999,Dobaczewski:1996,Tajima:2004,
Tichai:2019,Robin:2021,Hoppe:2021,Fasano:2022,Chen:2022,Hu:2022,Hagen:2022,
Kortelainen:2022,Tichai:2022}. Unlike the sets
of natural orbitals used in these papers, and many similar studies in atomic physics and chemistry calculations, 
the sets we discussed here are accurate, have  
no negative canonical occupation probabilities, as they should, can be generated easily with the  
expected spherical symmetry, and their quality can be easily improved during 
the calculations by adapting the properties of the potentials $V, V_{so}$, $\mu$, and $\Delta$ 
to ensure high accuracy within a relatively small size basis set. A particular aspect which we observed
is that the canonical wave functions $\phi_k(\xi)$ depend very weakly
on the magnitude of the pairing field $\Delta$.  

Induced nuclear fission and collisions of heavy-ions are particularly relevant highly non-equilibrium strongly interacting quantum 
many-body system to study. Nuclear fragments emerge highly excited in both fission and heavy-ion collisions, 
with an average temperature well above the critical temperature $T>T_c$~\cite{Bulgac:2016,Bulgac:2019c,Bulgac:2022}, at
an excitation energy at which the pairing correlations are absent. Therefore, the fact that we 
formally obtained the result that the final quantum Boltzmann 
one-body/entanglement  entropy increases 
within a formalism emerging from a treatment of pairing correlations, in particular even after performing a projection 
on the total proton and neutron numbers, furthermore underlines our conclusion that this increase is indeed solely related to 
the significant larger degree of complexity and more entanglement in 
the final many-body wave-function, compared to the initial many-body function.  

Long tails of the momentum distribution
have been measured~\cite{Hen:2014,Hen:2017} and 
a comprehensive picture of nuclei should and can incorporate both mean field and 
SRCs~\cite{Bulgac:2022,Bulgac:2022a}. These conclusions are in full agreement with Shina 
Tan's~\cite{Tan:2008a,Tan:2008b,Tan:2008c,Zwerger:2011,Braaten:2011,Castin:2011} conclusion 
that the presence of SRCs, although not necessarily always with ``pure'' character $n_k=C/k^4$, 
are present irrespective of whether the system is superfluid or a Fermi-like liquid. 
The presence of these long tails for $n_k$ leads 
to a generalization of the textbook definition~\cite{Abrikosov:1963} of the equilibrium single-particle occupation probabilities 
in strongly interacting many fermion systems~\cite{Bulgac:2022a}. 

As the example of induced fission shows,
the current implementation of the extension to superfluid systems of the Time-Dependent DFT (TDDFT) includes 
single-particle momenta up to $\approx$ 600 MeV/c, the upper limit considered in current implementations of the chiral 
Effective Field Theory for nucleon interactions in the treatment of light, medium and even heavy nuclei. 
Upon including the proton-neutron dynamical pairing correlations one would be able to basically
describe, within a unified approach, both long-range and short-range nucleon correlations~\cite{Bulgac:2022}, 
particularly for non-equilibrium processes.   The highly non-equilibrium nuclear fission process discussed here is  apparently  
the largest system where quantum entanglement has been studied so 
far~\cite{Milburn:1997,Vidal:2003,Korepin:2004,Kitaev:2006,Levin:2006,Li:2008,
Chuchem:2010,Pal:2010,Bardarson:2012,Cohen:2016,Abanin:2019,Sinha:2020,Wimberger:2021, Liu:2022,Mueller:2022,Schneider:2022}, 
with aspects related to the widely studied topics of Hilbert space and many-body localization.

As a result of our present analysis we expect that the properties 
of the canonical basis set and the use of entanglement entropy
can be extended to strongly correlated quantum many-body 
system in order to characterize the degree of complexity of the corresponding many-body 
wave functions and the degree of their entanglement, and thus provide additional 
insight into the QIS of many-body systems and their dynamics. As a side result, 
we provided a method to construct easily a set of approximate canonical wave 
functions/natural orbitals with correct quantum numbers,  and which can be 
improved while a solution to the many-body problem is constructed.

 \vspace{0.5cm}

 The funding 
from the US DOE, Office of Science, Grant No. DE-FG02-97ER41014 and
also the support provided in part by NNSA cooperative Agreement
DE-NA0003841 is greatly appreciated. 
This research used resources of the Oak Ridge
Leadership Computing Facility, which is a U.S. DOE Office of Science
User Facility supported under Contract No. DE-AC05-00OR22725.

\appendix*\section{A}\label{sec:A}
 
We will review here the Bogoliubov-Valatin formalism and the definition of the canonical wave functions. 
As Klich~\cite{Klich:2006} has shown, the same formalism 
can be used to characterize the entanglement entropy of any non-interacting system partitioned into two complementary parts.
At the same time the reader should be aware that the definition of the entanglement entropy is not unique, as there are many
different way to partition a system into two subsystems and we refer the reader to reviews where these differences have been discussed
in rather great detail~\cite{Amico:2008,Eisert:2010,Horodecki:2009,Haque:2009}.

The creation $\alpha^\dagger_k$ and annihilation $\alpha_k$ quasiparticle operators are represented with 
a unitary transformation from the field operators as follows~\cite{Ring:2004}
\begin{align} 
\!\!\!\!\!\! &\alpha_k^\dagger  =  
\int d\xi\left [ {\textrm u}_k(\xi) \psi^\dagger (\xi) + {\textrm v}_k(\xi) \psi (\xi)\right ], \label{eq:a0}\\
\!\!\!\!\!\! &\alpha_k= 
\int d\xi\left [ {\textrm v}_k^*(\xi) \psi^\dagger (\xi ) + {\textrm u}_k^*(\xi) \psi (\xi)\right ], \label{eq:b0}
\end{align}
and with the reverse relations  
\begin{align} 
&\psi^\dagger (\xi) = \sumint_k \left [ {\textrm u}^*_k(\xi)  \alpha^\dagger _k  
                                            + {\textrm v}_k(\xi)\alpha_k \right ], \label{eq:p1}\\
&\psi(\xi) =                \sumint_k \left [ {\textrm v}^*_k(\xi)\alpha^\dagger_k
                                            + {\textrm u}_k(\xi)\alpha_k \right ]. \label{eq:p2} 
\end{align}
Here $\psi^\dagger (\xi)$ and $ \psi (\xi)$ are the field operators 
for the creation and annihilation of a particle with coordinate $\xi=({\bm r},\sigma,\tau)$, 
$(u_k(\xi),v_k(\xi))^{T}$ are the quasiparticle wave functions,
and the integral implies also a summation over discrete variables when appropriate.
In a finite volume, with periodic boundary conditions,  the index $k$ is always discrete. 
For a finite isolated system in vacuum~\cite{Bulgac:1980,Belyaev:1984}
the sum over $k$ stands for a summation over the discrete indices and an
integral over the continuous ones respectively.

The Hermitian number density and the skew-symmetric anomalous density matrices are defined as
\begin{align}  
&n(\xi,\zeta) = \langle \Phi|\psi^\dagger(\zeta)\psi(\xi)|\Phi\rangle  = \sumint_k {\textrm v}_k^*(\xi) {\textrm v}_k(\zeta), \label{eq:number_v}\\
&\overline{n}(\xi,\zeta) = \langle \Phi|\psi(\xi)\psi^\dagger(\zeta)|\Phi\rangle  = \sumint_k {\textrm u}_k(\xi) {\textrm u}^*_k(\zeta),\label{eq:number_u}\\
&\kappa(\xi,\zeta) = \langle \Phi|\psi(\zeta)\psi(\xi)|\Phi\rangle =\sumint_k {\textrm v}_k^*(\xi){\textrm u}_k(\zeta),\label{eq:number_a} \\
& n(\xi,\zeta)+\overline{n}(\xi,\zeta) = \delta(\xi-\zeta), \label{eq:ntr}
\end{align}
where the quasiparticle vacuum is defined as
\begin{align}
& \alpha_k|\Phi\rangle =0, \, |\Phi\rangle = {\cal N} \prod_k\alpha_k|0\rangle , \, \langle \Phi|\alpha_k\alpha_l^\dagger|\Phi\rangle =\delta_{kl},
\end{align}
and ${\cal N}$ is a normalization factor (determined up to an arbitrary phase), $\alpha_k|0\rangle \neq 0$, 
and $|0\rangle$ is the vacuum state.
For any $k$, if the norm $\int d\xi  |{\textrm v}_k(\xi)|^2 = 0$ the corresponding factor ${\alpha}_k$  
should be skipped in the definition of $|\Phi\rangle$.
The new density matrix $\overline{n}(\xi,\zeta)$ is used in the discussion of the canonical basis set. \footnote{The
wave functions $v_k(\xi)$ can be considered as either as the vectors labeled by $k$ with components enumerated by $\xi$ or as the 
vectors label by $\xi$ and components enumerated by $k$. Thus $n(\xi,\zeta)$ 
is the complex scalar product of vectors with labels $\xi$ and $\zeta$, while $\langle v_k|v_l\rangle$ is the complex scalar product of vectors $k$ and $l$.}  

The anti-commutation relations for the field operators $\psi^\dagger (\xi),\,  \psi (\xi)$ and for the quasiparticle operators $\alpha_k^\dagger, \alpha_k$
imply that~\cite{Ring:2004}  
\begin{align} 
&\int d\xi \, [ {\textrm u}_k^*(\xi){\textrm u}_l(\xi)  +  {\textrm v}_k^*(\xi){\textrm v}_l(\xi) ]= \delta_{kl},\label{eq:orthog}\\
&\int d\xi \, [ {\textrm u}_k(\xi)   {\textrm v}_l(\xi)  +  {\textrm v}_k(\xi)   {\textrm u}_l(\xi) ]= 0, \label{eq:uu_vv}\\ 
&\sumint_k \, [ {\textrm u}_k(\xi){\textrm u}^*_k(\zeta)  +  {\textrm v}^*_k(\xi){\textrm v}_k(\zeta)] =\delta(\xi-\zeta),\label{eq:ov_com}\\
&\sumint_k \, [ {\textrm u}_k(\xi){\textrm v}_k^*(\zeta)  +  {\textrm v}_k^*(\xi){\textrm u}_k(\zeta) ]=0.
\end{align}
Eq.~\eqref{eq:ov_com} means that the quasiparticle wave functions ${\textrm u}_k(\xi), \,{\textrm v}_k(\xi)$ 
form (in general) an over complete set, as
for an arbitrary function $g(\xi)$ one has the decomposition 
\begin{align} 
 g(\xi) = &\sumint_ k\, {\textrm u}_k(\xi) \int d\zeta \, {\textrm u}_k^*(\zeta)g(\zeta) \nonumber \\
 +&   \sumint_k \, {\textrm v}_k^*(\xi) \int d\zeta \, {\textrm v}_k(\zeta)g(\zeta) .
 \end{align}
Additionally one can show that~\cite{Ring:2004}
\begin{align}
&\int d\zeta \, [ n(\xi,\zeta)n(\zeta,\eta) + \kappa(\xi,\zeta)\kappa^\dagger(\zeta,\eta) ] = n(\xi,\eta), \label{eq:nn}\\
& \int d\zeta \, n(\xi,\zeta)\kappa(\zeta,\eta) = \int d\zeta \, \kappa(\xi,\zeta)n^*(\zeta,\eta). \label{eq:skew}
\end{align}

For a finite system the quasiparticle components ${\textrm v}_k(\xi) $ 
always have a finite norm~\cite{Bulgac:1980}
\begin{align} 
\int d\xi \, |{\textrm v}_k(\xi)|^2 <\infty,
\end{align}
unlike the quasiparticle components ${\textrm u}_k(\xi)$, which can be either normalizable or not in an infinite volume. 
The index $k$ can be either discrete or continuous respectively. 
 
One can consider an arbitrary unitary transformation ${\cal U}{\cal U}^\dagger =\mathbb{I}$ (where $\mathbb{I}$ is the identity operator) 
of the quasiparticle wave functions 
\begin{align}
& \tilde{{\textrm v}}_l =\sumint_{k} {\cal U}_{kl} {\textrm v}_k, \quad  {\textrm v}_k =\sumint_{l} {\cal U}^*_{kl} \tilde{{\textrm v}}_l, \\
& \tilde{{\textrm u}}_l =\sumint_{k} {\cal U}_{kl} {\textrm u}_k, \quad  {\textrm u}_k =\sumint_{l} {\cal U}^*_{kl} \tilde{{\textrm u}}_l,
\end{align} 
which leaves the normal and anomalous density matrices unchanged. This type of transformation for quasiparticle wave functions was suggested in Refs.~\cite{Bulgac:2019a,Bulgac:2021c} in order to simultaneously diagonalize the 
overlap matrices $\langle {\textrm v}_k | {\textrm v}_l \rangle$ and  $\langle {\textrm u}_k | {\textrm u}_l \rangle$.
Only the canonical occupation probabilities are invariant with 
respect to arbitrary unitary transformations ${\cal U}$ mentioned above and 
one can then show that
\begin{align} 
&n(\xi,\zeta) = \sumint_k n_k\phi_k^*(\xi)\phi_k(\zeta), \label{eq:norm}\\
&\kappa(\xi,\zeta)= \sumint_k \sqrt{n_k(1-n_k)} \phi_{\overline{k}}^*(\xi)\phi_k(\zeta),\label{eq:anom}
\end{align}
where 
\begin{align}
&\phi^*_{\overline{k}}(\xi) = \frac{1}{\sqrt{n_k(1-n_k)}}\int d\zeta \,\kappa(\xi,\zeta)\phi_k^*(\zeta),\label{eq:kbar}\\
& \langle \phi_k|\phi^*_{\overline{k}}\rangle =0, \quad \langle \phi^*_{\overline{k}}|\phi^*_{\overline{k}}\rangle =1,\label{eq:e1}\\
&\int d\zeta \, n(\xi,\zeta)\phi^*_{\overline{k}}(\zeta) = n_k\phi^*_{\overline{k}}(\xi),\label{eq:e2}
\end{align}
where only $0<n_k<1$ contribute in Eqs.~(\ref{eq:anom}, \ref{eq:kbar}) and Eqs.~(\ref{eq:e1}, \ref{eq:e2}) 
follow from Eqs.~(\ref{eq:nnn1}, \ref{eq:nn}, \ref{eq:skew}). 

In the Hartree-Fock (HF) approximation the situation is much simpler, since $n_k=0$ or 1, the anomalous density vanishes and
the occupation probabilities are defined in the representation which simultaneously diagonalizes the number density matrix  
and the mean field, and in that particular representation the occupation probabilities have a straightforward physical interpretation. In the presence of 
pairing correlations one can introduce a generalized density matrix~\cite{Ring:2004}, 
which commutes with the generalized mean field. 
However, in that representation the normal number density has the form given by Eq.~\eqref{eq:number_v}, where,
$\langle {\textrm v}_k|{\textrm v}_l\rangle \neq n_k\delta_{kl}$. One can define the occupation probabilities either as 
$n_k= \langle {\textrm v}_k|{\textrm v}_k\rangle$ in the representation in which the generalized mean field is diagonal, or instead 
use the canonical  occupation probabilities $n_k$ from Eq.~\eqref{eq:nnn1} and define the single particle energies as $e_k= \langle \phi_k|H|\phi_k\rangle$, 
where $H$ is the normal mean field single-particle Hamiltonian within the Hartree-Fock-Bogoliubov (HFB) 
and Superfluid Local Density Approximation (SLDA) frameworks, and in which case $\langle \phi_k|H|\phi_l\rangle \neq 0$ if $k\neq l$. 
The simple relationship between the HF occupation 
probabilities and the single-particle energies thus becomes more difficult to interpret physically and justify within HFB 
and SLDA frameworks. 
\footnote{For any
many-fermion system, either superfluid or normal, can be described using the same formalism, as one can introduce
a normal number density matrix Eq.~\eqref{eq:norm} and an ``anomalous number density'' Eq.~\eqref{eq:anom} with the functions 
$v_k(\xi)=\sqrt{n_k}\phi_k(\xi)$ and $u_k(\xi) = \sqrt{1-n_k}\phi_{\overline{k}}(\xi)$~\cite{Klich:2006}.} 

Since the total particle number is not well defined within HFB and SLDA, as the gauge symmetry is broken, 
one has to restore this symmetry. In the canonical representation the gauge symmetry is significantly easier 
to restore~\cite{Bulgac:2019a,Bulgac:2021c}. The many-body wave-function acquires the well- known BCS form 
$| \Phi \rangle = \Pi_k(u_k+v_k a^\dagger_k a^\dagger_{\overline{k}}) | 0 \rangle$~\cite{Bardeen:1957}, 
where $a_k|0\rangle =0$,   $u^2_k+v^2_k=1$, and $n_k = v_k^2$  and various other gauge symmetry 
restored observables can be easily extracted~\cite{Bulgac:2021c}.  

The quasiparticle representation in which the generalized number density matrix and the generalized mean field commute is particularly 
suited for  numerically determining the corresponding static many-body wave-function $|\Phi\rangle$, only if one uses diagonalization methods. 
The diagonalization, which is typically numericaly very expensive, can be eschewed~\cite{Jin:2017}, as 
both normal and anomalous densities can be determined without the 
knowledge of the quasiparticle wave functions (qpwfs) and of the corresponding quasiparticle energies.

\providecommand{\selectlanguage}[1]{}
\renewcommand{\selectlanguage}[1]{}

\bibliography{local_fission}

\end{document}